\documentclass{aa}  

\usepackage[varg]{txfonts}
\usepackage{multirow}
\usepackage{natbib}
\bibpunct{(}{)}{;}{a}{}{,} % to follow the A&A style
\usepackage{xcolor}
\usepackage{breqn}
\usepackage[colorlinks=true]{hyperref}
\usepackage[xindy]{glossaries}
\hypersetup{linkcolor=red,citecolor=blue,filecolor=cyan,urlcolor=green}
   
\usepackage{graphicx}

\newcommand*\subtxt[1]{_{\mathrm{#1}}}
\DeclareRobustCommand\_{\ifmmode\expandafter\subtxt\else\textunderscore\fi}

\DeclareMathAlphabet{\mathcal}{OMS}{cmsy}{m}{n}
\SetMathAlphabet{\mathcal}{bold}{OMS}{cmsy}{b}{n}
\definecolor{RedOrange}{HTML}{F26035}

\definecolor{ForestGreen}{HTML}{00FF00}

\definecolor{Azure}{HTML}{007FFF}

\begin{document} 

   \title{Properties of loss cone stars in a cosmological galaxy merger remnant}

   \author{Branislav Avramov\inst{1}, 
           Peter Berczik \inst{2, 1, 3}, 
           Yohai Meiron \inst{4, 5}, 
           Anshuman Acharya\inst{6} 
          \and
          Andreas Just\inst{1}
          }

   \institute{Astronomisches Rechen-Institut (ARI), Zentrum f\"ur Astronomie, University of Heidelberg, M\"onchhofstr. 12-14, 69120 Heidelberg, Germany  \\
         \email{branislav.m.avramov@gmail.com}
         \and    
         National Astronomical Observatories and Key Laboratory of Computational Astrophysics, Chinese Academy of Sciences, 
         20A Datun Rd., Chaoyang District, Beijing 100101, China
         \and  
         Main Astronomical Observatory, National Academy of Sciences of Ukraine, 27 Akademika Zabolotnoho St., 03143 Kyiv, Ukraine 
        \and 
        Department of Astronomy and Astrophysics, University of Toronto, 
        50 St. George Street, Toronto, ON M5S\,3H4, Canada
        \and 
        SciNet High Performance Computing Consortium, University of 
        Toronto, 661 University Ave., Toronto, ON M5G\,1M1, Canada
        \and 
        Indian Institute of Science Education and Research (IISER) - Mohali, SAS Nagar, Punjab, 140306, India. 
        }

   \date{Received 16 October 2020; accepted 26 February 2021}

% \abstract{}{}{}{}{} 
% 5 {} token are mandatory
 
  \abstract
   {}
   {We investigate the orbital and phase space properties of loss cone stars that interact strongly with a hard, high-redshift binary supermassive black hole (SMBH) system formed in a cosmological scenario. }
   {We use a novel hybrid integration approach that combines the direct N-body code $\varphi$-GRAPE with ETICS, a collisionless code that employs the self-consistent field method for force calculation. The hybrid approach shows considerable speed-up over direct summation for particle numbers $> 10^6$, while retaining accuracy of direct N-body for a subset of particles.  During the SMBH binary evolution we monitor individual stellar interactions with the binary in order to identify stars that noticeably contribute to the SMBH binary hardening.}
   {  We successfully identify and analyze in detail the properties of stars that extract energy from the binary. We find that the summed energy changes seen in these stars match very well with the overall binary energy change, demonstrating that stellar interactions are the primary drivers of SMBH binary hardening in triaxial, gas-poor systems. We find that $76\%$ of these stars originate from centrophilic orbits, only possible in a triaxial system. As a result, even the slight triaxiality of our system results in efficient refilling of the loss cone,  avoiding the final parsec problem. We distinguish three different populations of interactions based on their apocenter. We find a clear prevalence of interactions co-rotating with the binary. Nevertheless, retrograde interactions are the most energetic, contributing only slightly less than the prograde population to the overall energy exchange. The most energetic interactions are also likely to result in a change of sign in the angular momentum of the star.  We estimate the merger timescale of the binary to be $\approx 20$ $\mathrm{Myr}$, a value larger by a factor of two than the timescale reported in a previous study. }
   {}


\titlerunning{Properties of Loss Cone Stars}
\authorrunning{B. Avramov et al.}
\maketitle
%
%-------------------------------------------------------------------

\section{Introduction}

Supermassive black holes (hereafter SMBHs) are present in the centers of the vast majority of galaxies \citep[see e.g.,][]{Magorrian1998}. During a major galaxy merger,  SMBHs are expected to be able to form a close, Keplerian binary pair which would merge at a later stage, followed by gravitational wave emission. The evolution of an SMBH binary was first described in \citet{Begelman1980}, and since then has been the subject of numerous studies \citep[for a review see][]{Colpi_2014}.

So far, detections of SMBH binaries remain largely elusive, despite large observational efforts \citep{DeRosa2019, Bogdanovic2015}. However, several promising candidates have been identified using very long baseline interferometers  \citep[VLBIs, ][]{Rodriguez2006, Kharb2017}, or with photometric measurements \citep{Liu2014, Valtonen2008}. More recently, a triple SMBH system was observed with MUSE \citep{Kollatschny2020}.

The three distinct phases characterizing the evolution of an SMBH pair are the dynamical friction phase,  stellar hardening phase, and gravitational wave (GW) inspiral \citep{Begelman1980}. \textit{Dynamical friction phase:}  Owing to dynamical friction with the surrounding stars, the SMBHs sink towards the center of the potential of the galaxy merger remnant  \citep[GMR;][]{Chandra_1943, Just2011}. As the separation between the black holes decreases,  SMBHs eventually become close enough to form a binary system. At this point, until the binary becomes hard, dynamical friction effects resulting from many small-angle encounters start to become less relevant, while individual large-angle scatterings with stars start to play a more prominent role. \textit{Stellar hardening phase:}  In this phase of the binary evolution,  the primary mechanism of energy loss for the binary is via close interactions of stars, referred to as slingshot ejections. Namely, incoming stars  extract orbital energy from the binary via three-body interactions, thus increasing their own kinetic energy. \textit{Gravitational wave inspiral phase:}  At separations of $10^{-3}\text{-}10^{-2}$ pc, energy loss due to GW emission becomes significant and brings the binary to coalescence on a well-defined timescale \citep{PetersMathews63,Zwick2020, Zwick2021}. These events are expected to be the loudest gravitational wave sources for present and upcoming gravitational wave searches, namely  the  PTA (Pulsar Timing Array) search for SMBH binary systems with $M\gtrsim
10^8$ $\mathrm{M_\odot}$ \citep{Mingarelli2017, Kelley2018} and the upcoming LISA (Laser Interferometer Space Antenna) mission \citep{LISA2017} for SMBH binaries in the $10^4\text{-}10^7$ $\mathrm{M_\odot}$ range.   

Stars that can potentially interact strongly with the SMBH binary come from a region in phase space known as the loss cone \citep{Yu2002, Milosavljevic2003}. A star from the loss cone has sufficiently low angular momentum to approach the binary at a distance of $\sim 2a\_{bh}$, where $a\_{bh}$ is the SMBH binary semi-major axis. It can then either be tidally disrupted by one of the black holes, ejected from the surrounding environment (while staying bound to the system, and therefore possibly interacting once more in the future), or ejected from the system entirely with a velocity $v> v\_{esc} \sim 1000$ $\mathrm{[km\, s^{-1}]}$ as a hypervelocity star (HVS). The existence of HVSs was first proposed by \citet{Hills1988}, as a result of a close encounter between a stellar binary and an SMBH. \citet{Yu2003} also explored a scenario where an SMBH binary produces HVSs via scattering processes.  

\citet{Quinlan1996} established the expected hardening rates of an SMBH binary owing to stellar interactions. Since then, numerous studies using isolated scattering experiments have been performed in order to shed light on how these interactions affect the properties of a massive black hole binary \citep[e.g.,][]{Bonetti2020,Sesana2006,Sesana2008, Rasskazov2019HVS}.  The main benefit of the scattering experiments is  their low computational cost, and therefore their ability to estimate the properties of a large number of encounters over many different SMBH binary configurations (e.g., different mass ranges, eccentricities, separations, etc.). 

However, large-scale N-body simulations are able to more robustly estimate SMBH merger timescales. According to these,  efficient hardening of an SMBH binary is necessary for merger timescales that are shorter than a Hubble time, making any future GW detections by LISA reliant on the assumption that there are enough stars in the loss cone to interact with the binary. Studies show that the ejection of loss cone stars by the binary results in a gap in the loss cone phase space \citep{Merritt2005}. If this were true, within a dynamical timescale, the loss cone would be depleted and it would lead to stalling of the SMBH binary, unless the loss cone can be efficiently repopulated. This can occur via collisional processes, where stellar two-body relaxation decreases a star's angular momentum and scatters it into the loss cone. However, galaxies, unlike globular clusters, are largely collisionless systems, and the relaxation timescale for a galaxy is larger than the Hubble time \citep{Berczik2005, Milosavljevic2001}. This issue has led to the so-called  ``final parsec problem''  \citep[FPP,][]{Milosavljevic2003}. 
 
Another mechanism of loss-cone refilling is via collisionless processes. More specifically, in nonspherical systems, torques from the stellar potential can decrease stellar angular momentum in such a way that the star is put on a loss cone orbit \citep{Merritt_Poon2004}. Gravitational two-body relaxation may still play a role, albeit on a much longer timescale.  Nonspherical features result in distinct orbital families, some of which are centrophilic. In axisymmetric systems, orbits have conservation of the z-component of angular momentum, which represents a lower boundary to their total angular momentum.  Therefore,  stars can enter the loss cone and interact with the binary only when their z-component of angular momentum fulfils the following relation: $L_z \leq \sqrt{2GM\_{BH}a\_{BH}}$,  where $G$ is the gravitational constant, $M\_{BH}$ the total SMBH binary mass, and $a\_{BH}$ the binary semi-major axis. These orbits are often referred to as  saucer  orbits \citep{Merritt2013}. In triaxial systems, along with the saucers, there is also another family of loss-cone orbits, the pyramid orbits \citep{Merritt2013, Merritt2011}, which can be seen as analogs of regular box orbits in triaxial potentials \citep{Binney1987}. The main difference between these two orbital types is  that in the case of pyramids, there is an SMBH which serves as a reflecting boundary \citep{Merritt2011}. The stars on these orbits can attain an arbitrarily small value of angular momentum, where the only constraint is $L\_{min}>0$, and are therefore truly centrophilic in nature.  With the growing consensus that observed galactic nuclei have nonspherical and triaxial features, 
collisionless processes seem like a natural and consistent solution to the problem of efficient loss-cone refilling. 

The FPP arose from modeling idealized spherical systems, leading to the conclusion that on its own, two-body relaxation effects are not enough to refill the loss cone within a Hubble timescale \citep[][]{Berczik2005}.  Since then,  N-body studies have shown that triaxial systems, even those which deviate only slightly from spherical symmetry, result in efficient refilling of the loss cone and hardening rates that are sufficient to bring the binary to coalescence,  thus resolving the FPP \citep{Gualandris2017, Vasiliev_2015,Preto2011,Khan2011, Berczik2006}. However, in the case of axisymmetric galaxies, the picture is less clear. A solution to the FPP  was reported in the axisymmetric case for galaxies in equilibrium  \citep{Khan2013},  as well in rotating galaxies \citep{Khan2020, Mirza2017, Holley-Bockelmann2015}. On the other hand, \citet{Vasiliev_2014} find that axisymmetry alone is not enough, but that even small deviations from axisymmetry can lead to coalescence.
Additionally, studies report that the presence of massive stellar perturbers or stellar clusters \citep{ArcaSedda2019, Bortolas2018}, a gaseous circumbinary  \citep{Escala2005, Cuadra2009, Roedig2014} or circumnuclear \citep{SouzaLima2020} disk,  gaseous clumps and molecular clouds \citep{Goicovic2017, Goicovic2018}, as well as subsequent galaxy mergers and black hole triplets \citep{Iwasawa2006, Tanikawa2011, Bonetti2018, Ryu2018} can also accelerate SMBH coalescence and therefore help avoid the FPP.  Brownian motion of the SMBH binary can also result in additional diffusion of stars in the loss cone \citep{Bortolas2016}. Furthermore, the presence of a rotating stellar cusp can affect the SMBH binary properties, as well as its interactions with incoming stars \citep{Rasskazov2017,Wang2014,Li2012}. Most recently, \citet{Ogiya2020} showed that tidal effects from merging nuclear star clusters can also significantly shorten the SMBH merger timescale. As a result of efficient collisionless loss cone refilling, as well as other effects listed above, merger timescales are now estimated to be $t\_{merger} \lesssim 1$ $\mathrm{Gyr}$ \citep{Biava2019, Khan2018a, Khan2018b, Rantala2017}.

\par However, cosmological context and a cosmological environment can be crucial factors that affect galaxy morphology and dynamics in a merger and therefore determine the orbital nature of loss-cone stars and the evolution of the SMBH binary system. Recently, several studies have explored the conditions for the formation of an SMBH pair in a cosmological context \citep{Bortolas2020, Pfister2019, Tremmel2018-2, Tremmel2018}, showing that along with dynamical friction, additional  more complex effects might be in play in the early evolution of an SMBH binary. Nevertheless, \citet{Khan2016} remains the only example in the literature so far where the entire evolution of an SMBH binary is modeled, starting from a cosmologically formed galaxy merger, down to the post-Newtonian (PN) plunge of the black holes themselves.  In that study,   the resulting GMR was slightly triaxial with very high central stellar densities, which led to a constant hardening rate resulting in the merger of the black holes in just $10$ $\mathrm{Myr}$ after the binary formation. This latter study was therefore in a unique position to explore the properties of the orbits of the stars which contributed to the binary hardening and to investigate loss-cone-refilling mechanisms originating from a robust, cosmological environment. However, as in many direct N-body studies that came before, the authors fell short of exploring this aspect. 

The reason for this lies in the nature of the standard N-body approach.  Along with high computational cost, one of the main drawbacks of N-body is that, because the number of particles that can be simulated is significantly smaller than a realistic galaxy, nonphysically massive particles may cause artificially enhanced two-body relaxation effects, thus unintentionally increasing the effect of collisional hardening processes and making it inherently difficult to consistently study mechanisms of loss-cone refilling. It should also be noted that in a similar fashion,  Brownian motion of a binary in an N-body system can be enhanced because of insufficient particle numbers and can artificially increase the hardening rate \citep{Milosavljevic2003-2}, but this effect is not expected to be significant in simulations with $N > 10^6$  particles \citep{Bortolas2016}. Therefore, to circumvent the issue, a sort of hybrid approach is needed,  which would retain the accuracy of direct summation for particles interacting with the SMBH binary, but would be able to use collisionless expansion techniques (or sufficiently large particle numbers) to accurately model the outer regions of the system. Recently, several similar approaches using different methodologies have been proposed in the literature. \citet{Mannerkoski2019} and \citet{Rantala2018} use a hybrid-tree code \citep{Rantala2017}, combining GADGET-3 \citep{Springel2005} with algorithmic regularization \citep{Mikkola2008} to investigate SMBH merger timescales. \citet{Lezhnin2019} used a Monte-Carlo approach \citep{Vasiliev_2015} to investigate tidal disruption rates and ejection of stars from the loss cone. Finally, \citet{Nasim2020} and \citet{ Gualandris2017} used a Fast Multiple Method code  combined with direct summation to simulate SMBH binary evolution.

 In the present paper, we perform a detailed study of individual stellar hardening interactions with a hard SMBH binary in a high-redshift galactic merger.  We  do this by resimulating the final stages of the merger in the \citet{Khan2016} system, the only example in the literature of an SMBH binary followed from cosmological origin to coalescence. We use $\varphi$-GRAPE-hybrid (Meiron et al. 2021, in prep.),  a novel hybrid integration approach that combines the direct N-body code $\varphi$-GRAPE \citep{Harfst2007} with ETICS \citep{Meiron2014}, a self-consistent field method (SCF) code.  In this way, we ensure sufficient integration accuracy for particles of interest, while avoiding the high computational cost and enhanced relaxation effects that typically plague pure N-body approaches.  We include  PN effects in the equations of motion up to order 2.5 \citep{Blanchet2006} in order to account for GW emission. 

 The paper is organized as follows. We start by specifying the numerical details of the codes and initial conditions in Section \ref{sec:Code}. In Section \ref{sec:results} we present our results. More specifically, in Section \ref{sec:bhs}, we present an overview of the hardening rates and orbital properties of the SMBH binary. In Section \ref{sec:GMR} we explore the shape and rotational properties of the galaxy merger remnant. Section \ref{sec:loss_cone} contains our main findings, where we analyze the orbits and properties of stars in the loss cone.  In Section \ref{sec:disc} we discuss our findings. Finally, in Section \ref{sec:conc}, we make a brief summary and present our main conclusions.

\section{Numerical description and initial conditions}
\label{sec:Code}
\subsection{Numerical methods}

For our simulation we used the $\varphi$-GRAPE-hybrid code (Meiron et al. 2021, in prep.), an adapted hybrid version of the direct N-body code $\varphi$-GRAPE \citep{Harfst2007}. The main attribute of $\varphi$-GRAPE-hybrid is that it combines two different force calculation methods, the self-consistent field method \citep[hereafter SCF,][]{Hernquist1992} with direct summation. This is done through a special library called GRAPite, which serves as a bridge between $\varphi$-GRAPE and ETICS \citep{Meiron2014}, a SCF code. The SCF method is a multipole expansion method used for collisionless systems which solves the Poisson equation by decomposing the potential into three series of multipoles:  
\begin{equation}
\label{eq:scf}
\Phi (r) =  \sum_{n=0}^{\infty}\sum_{l=0}^{\infty}\sum_{m=-l}^{+l}A_{nlm}\Phi _{nl}(r)Y_{lm}(\theta, \varphi), 
\end{equation}
where $A_{nlm}$ are the expansion coefficients, $Y_{lm}$ are the spherical harmonics, and $\Phi _{nl}$ are the radial basis functions. ETICS adopts the radial basis set of \citet{Hernquist1992}, where the zeroth-order ($n, l, m= 0$) is a Hernquist profile  and the subsequent orders of $\Phi_{nl} (r)$ are calculated using Gegenbauer polynomials. The two infinite series in Eq.  \ref{eq:scf} need to be cut off at $(n\_{max}, l\_{max})$, where $n\_{max}$ corresponds to the maximum radial expansion order and $l\_{max}$ depends on the deviation of the system from spherical symmetry. For our simulation, we adopt the standard choice of $(n\_{max}, l\_{max}) = (10, 6)$ \citep{Meiron2014}. 

The implementation of SCF through ETICS enables fast integration of a large number of particles, because not every interaction is calculated separately. However, due to the nature of the method, the SCF cannot account for two-body relaxation effects. These effects are irrelevant for large systems like galaxies where the relaxation timescale is much larger than the dynamical timescale, but become important when looking at the environment around an SMBH binary and the stellar interactions with it. On the other hand, direct summation integration is  able to accurately compute the gravitational interaction between an SMBH binary component and an interacting star, but at the cost of very large computation times. Therefore, the main benefit of a hybrid approach is the ability to select a subset of particles whose interactions would be computed directly, while the rest of the particles, with negligible two-body relaxation effects,  would be computed using the SCF method.  This approach significantly reduces computational time, while retaining integration accuracy for particles of interest. 
\par In $\varphi$-GRAPE-hybrid, this is accomplished by dividing the particles into three categories: \textit{core, halo,} and \textit{black holes}. It is important to note that, as opposed to what the name suggests, \textit{halo} particles do not necessarily belong to the galactic halo (similarly, \textit{core} particles do not necessarily belong to the galactic core). Instead, \textit{halo} particles is the collective name for the collisionless particles, while \textit{core} particles is the collective name for particles integrated with direct summation. In the code, \textit{core} particles have direct interactions amongst themselves, while \textit{halo} particles interact amongst themselves as well as with the \textit{core} particles through the SCF force expansion. Black hole particles have direct interactions with all particles in the system. The scaling of the force calculation with the number of particles is of the order $\mathcal{O}(N\_{core}^2)+\mathcal{O}(N\_{halo})$, assuming that the number of black hole particles is of order unity. If there is only a small fraction of \textit{core} particles, the speed-up obtained over the direct-only approach is very significant. 
\par The direct part of the force calculation   is handled by the Yebisu GRAPE emulation library \citep{Nitadori2009}. The code as a whole is a fourth-order Hermite integrator with block time-steps. Both direct and SCF force calculations are performed on GPUs, and the code is fully parallelized through MPI.

\subsection{Initial conditions}
\label{sec:ini}
As initial conditions for our simulations we use the same system as in \citet{Khan2016}.
In that work,  a massive galaxy merger at redshift $z\sim 3.5$ was identified and followed using the Argo cosmological simulation \citep{FeldmannMayer2015, Fiacconi_etal2015}. 
At a time which we refer to as the initial time $t\_{ini}$ throughout the text, a static particle-splitting procedure was performed in order to increase the particle number, and two SMBH particles with masses  $M\_{BH1} = 3 \times 10^{8} M_{\odot}$ and $M\_{BH2} = 8 \times 10^{7} M_{\odot}$ were introduced at the local minima of the gravitational potential of the galactic cores. 
    The system was evolved further using the GASOLINE code \citep{Wadsley2004} and during the final stages of the merger, the galaxy merger remnant had a gas fraction of only 5\%. At time $t\_{PN} = t\_{ini}+21.5$ $\mathrm{Myr,}$ the remaining gas particles were turned into star particles,  a spherical region of 5 kpc around the most massive SMBH was extracted, the softening was further reduced, and the PN terms were turned on. At this stage, the separation between the black holes was $\sim 300$ $\mathrm{pc}$.   This system was then further evolved using the direct N-body code $\varphi$-GPU \citep{Berczik2011}. Integration was continued until the merger of the SMBH particles was induced by the PN corrections (up to order 3.5) in the equations of motion. For more details on the simulation setup, we refer the reader to \citet{Khan2016}. 

In order to resimulate the final stages of the merger, we initialize our simulations at $t\_0=t\_{ini}+27.7$ $\mathrm{Myr}$ = $t\_{PN}+6.2$ $\mathrm{Myr}$, which we refer to as the resimulation time. At this time the SMBH binary is well into the hardening phase of the merger with no measurable PN effects and the binary semi-major axis  is $a\_{bh}\sim 0.06$ $\mathrm{pc}\sim 1650 R\_{sch}$, where $R\_{sch}=3.6\times10^{-5}$ $\mathrm{pc}$ is the Schwarzschild radius of the combined mass of the binary. As in the final N-body run in \citet{Khan2016}, our particle number is $N \sim 6 \times 10^6$ particles, consisting of two SMBH particles, 414 414 dark matter particles, and 5 511 152  star particles. The entire system has a mass of $9.18\times10^{10} M_{\odot}$, while the dark matter and typical stellar particle masses are $1\times10^5 M_{\odot}$ and $9.1\times10^{3} M_{\odot}$, respectively. 

\par The particles which are treated in a direct way (\textit{core} particles) need to be selected before the initialization of the run. In our system, while relaxation effects are negligible in the outer regions of the system,  they become important in the central region. We therefore choose \textit{core} particles to reflect this fact.  In order to estimate the region of influence of the SMBH binary, we  use the cumulative radial mass profile of the GMR. Namely, we define the influence radius of the SMBH binary as the radius at which the following condition is satisfied: 
\begin{equation}
 M\_{cum}(R\_{infl}) = 2M\_{BH},   
\end{equation}
 where $M\_{cum}(r)$ is the cumulative mass of all of the particles (excluding the SMBHs) within the sphere of radius $r$ and $M\_{BH}$ is the total mass of the SMBH binary $M\_{BH} = M\_{BH1}+M\_{BH2}$. This corresponds to the radius of $R\_{infl} = 13.15$ pc.
 In order to ensure that we select particles that have the potential to interact strongly with the binary, we determine stellar particles to be \textit{core} if they fulfill \textit{at least} one of the following criteria:

\begin{equation}
\label{eq:core_cond}
r\_p < 0.5 R\_{infl} \quad\text{or}\quad
r < R\_{infl},
\end{equation}

where $r\_{p}$ is the pericenter distance in the two-body Keplerian approximation of a star-BH binary system and $r$ is the radial distance of the particle at the start of the resimulation. This selection is performed only with stellar particles and only  once at the start of the run. Using this condition,  we obtain $~2 \times 10^{5}$ \textit{core} particles which will be calculated in a direct way. This corresponds to 3.3\% of our total particle number $N=6\times10^6$. 
We note however that as a result of the complex phase space structure of the GMR, there is a fraction of \textit{halo} stars that will enter the inner region at a later time, and possibly even become centrophilic. This fraction of stars is low ($\approx 18\%$ of all stars in the inner region are \textit{halo} at the end of the simulation) and has only a minor effect on the shape of the potential deep within the influence radius. These stars still have direct interaction with the black holes, and therefore can contribute to the overall hardening rate.  
 
\par The criterion in Eq. \ref{eq:core_cond} helps us to ensure that we represent the potential correctly at the time of resimulation in the transition between the Kepler-potential-dominated region inside $R\_{infl}$ and the outer region where the potential is dominated by the contribution of the GMR. Within the Kepler region, the potential of the system can be approximated to a high degree of accuracy by the  analytical expression:

\begin{equation}
\label{eq:pot}
\Phi(r < R\_{infl}) = -\frac{GM\_{BH}}{R\_{BH}}  + C, 
\end{equation}

where $C= const.$  approximates the contribution of the surrounding galactic system to the total potential within the Kepler region and $R\_{BH}$ is the distance to the SMBH binary center of mass. In order to estimate $C$, we perform a fitting procedure on the radial profile of the potential in the  centre of mass of the black holes reference frame. 
 The fitting is performed  at resimulation time $t\_0$ using the least squares method over a radial range of $R = [0.119,2.9]$ pc, with the zero-point of the potential being at infinity. We obtain the value of $C=-1.983 \times 10^6$ $\mathrm{km^2\,s}^{-2}$ 
 which corresponds to the minima of the residual of the fitting parameter. We find that within $0.3R\_{infl}$, Eq. \ref{eq:pot} does a good job of approximating the potential. Additionally, we do not observe measurable deviations in the potential over time. 
 \par In order to be able to identify and study the strong interactions between the stars and the SMBH binary, at every integration time-step we monitor each time a star enters and exits a sphere of radius $r = 10a\_{bh}$ around the black holes, where $a\_{bh}$ is the SMBH binary semi-major axis. This is done with additional, more frequent output of the black hole and interacting star paramaters. The time frequency of the additional output is  $1,500$ years, while the full snapshots are obtained every $12,000$ years, a factor of eight larger.
 This allows us to identify stars from the loss cone by analyzing the energy and orbital changes of the stellar particle before and after interaction, while eliminating any other effects that could cause such significant energy changes. 
 At this radius, the escape velocity is $V\_{esc}=3063$ $\mathrm{km\, s^{-1}}$, corresponding to the potential of $\phi = -4.692 \times 10^6$ $\mathrm{km^2\,s}^{-2}$. The SMBH binary gravitational pull constitutes about $58 \%$ of this value, while the effect of the GMR is represented by the remaining $42 \%$.

\subsection{Numerical parameters and performance}
\label{sec:num}
\par Because of the inherent difference between codes, as well as for increased accuracy, we change several simulation parameters from the setup used in \citet{Khan2016}. Most notably, we employ a global softening parameter of $\epsilon = 2\times10^{-4}$ $\mathrm{pc}$, while the original simulation had an individual softening approach depending on particle type. The smallest value of the softening was $\epsilon  = 0.007$ $\mathrm{pc}$, used for the black hole-star interactions. Our value is considerably smaller than the ones used in the original simulation, because we found that larger  values of the softening parameter  affected the hardening rate of the SMBH binary, resulting in artificially lower hardening rates.  Additionally, we considerably decrease the minimum black hole integration time-step in order to get better accuracy in the PN terms calculation.  The minimum integration time-step has  been decreased from $\Delta t\_{min} = 10^{-3}$ $\mathrm{yr}$ in the original run to $\Delta t\_{min} = 10^{-5}$ $\mathrm{yr}$ in our resimulation. This level of accuracy was previously unreachable for direct-only N-body approaches because of the already extremely high computational cost for a system with particle numbers $N > 10^6$.  The relative energy error throughout our run is typically $\Delta E/E \approx 10^{-5}$. However we record higher errors upon each  restart of the run (roughly every $0.8\, \mathrm{Myr}$), reaching  $\Delta E/E = 4.7\times 10^{-3}$. Therefore, for best results, frequent restarts should be avoided. 

\par Finally, our pilot runs showed that the choice of the origin of the reference frame can significantly affect the results and produce nonphysical hardening of the binary. This results from the fact that by design the SCF force is calculated at the origin of the reference frame. If the origin does not match with the density center of the system, the spatial asymmetry will introduce a bias in the SCF force, causing artificial sinking of the black holes towards the origin. However, in our system, the density center coincides with the SMBH binary center of mass, which has its own proper acceleration and therefore is a noninertial system. As an inertial reference frame is required in order to maintain conservation of energy and angular momentum,  using the SMBH binary center of mass as the origin is also not a desirable option. Therefore, to circumvent these issues,  we perform a coordinate frame change at the start of the run so that the origin matches with the position of the overall density center of the system at this time (which coincides with the black hole center of mass at this time). Additionally, we also assign a constant velocity to the origin equal to the mean velocity that the SMBH binary had in the original \citet{Khan2016} run in order to ensure our origin follows the mean movement of the SMBH binary throughout the run, and therefore also the density center.   In this way, we set up an inertial, comoving reference frame that follows the general movement of the density center and avoids unwanted bias from the SCF force calculation. Throughout the run, the SMBH center of mass undergoes a random walk  with respect to the origin. We check the amplitude of the drift of the black hole center of mass and find that the deviation  is neglible and does not exceed $25a\_{bh, 0}=1.5 \, \mathrm{pc}$, where $a\_{bh, 0}$ is the semi-major axis of the binary at resimulation time $t\_0$. This drift is too small to introduce bias in the SCF force and therefore the comoving reference frame setup resolves the above described issue successfully. Throughout the text, the results in the figures are centered to the SMBH binary center of mass, unless specified otherwise. 

\par We find that changes to the numerical parameters listed in this section affect the overall evolution of the SMBH binary, which we discuss in Section \ref{sec:bhs}.

\par Our benchmarks have shown that $\varphi$-GRAPE-hybrid has a speed-up of a factor of 16 over the direct-only version. This will be discussed in detail in an upcoming publication (Meiron et al. 2021, in prep).  Our case in particular is very well suited for a hybrid approach due to the very large range of scales that need to be resolved in order to obtain the proper physics of the system. These scales range from $1$ $\mathrm{mpc}$ for the close interactions to $10$ $\mathrm{kpc}$ for the outer ranges of the system. An even bigger performance issue is the range of timescales that are present in our system. The SMBH binary orbital period is on the order of $\sim1$ $\mathrm{yr}$, but the average crossing time of the system in on the order of $\sim1$ $\mathrm{Myr}$. Therefore, the time resolution needs to be high enough in order to properly resolve the three-body  interactions with incoming stars, which happen on the SMBH binary orbital timescale, while also evolving the surrounding system for several crossing times in order to allow for proper refilling of the loss cone. This obstacle comes at a great computational cost, even with the hybrid approach. In our case, the necessary decrease in the minimum integration time-step from $\Delta t\_{min} = 10^{-3}$ $\mathrm{yr}$  to $\Delta t\_{min} = 10^{-5}$ $\mathrm{yr}$ resulted in a $500 \%$ slowdown of the code.  However, despite this, the comparatively small percentage of particles calculated in a direct way and efficiency of the hybrid code enabled us to maintain sufficient speed on four computational nodes, each equipped with four Nvidia V100 GPU devices.

%--------------------------------------------------------------------

\section{Results}
\label{sec:results}

\subsection{SMBH binary properties}
\label{sec:bhs}

\par
\begin{figure*}
    \centering
    \includegraphics[width=\textwidth]{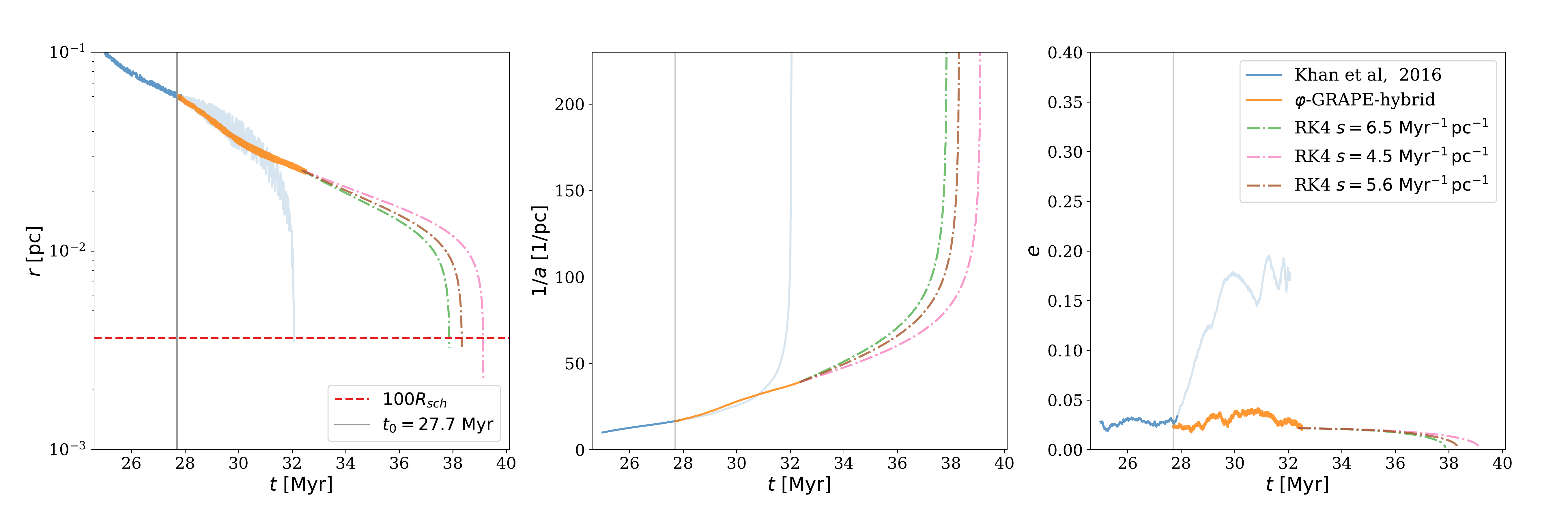}
    \caption{
    Left: Separation of the black holes as a function of time since $t\_{ini}$ (since the particle splitting procedure). The blue line corresponds to the original run in \citet{Khan2016}, while the orange line corresponds to our run. The light version of the blue line refers to the continuation of the original simulation data. The three dashed-dotted lines starting at $33.35$ $\mathrm{Myr}$ correspond to analytical estimates of the merger time in the PN-dominated regime using constant values of hardening rate. The vertical line marks the initial time of our run, $t\_0$. The horizontal dashed line marks the separation of $100R\_{sch}$, where the $R\_{sch}$ is the Schwarzschild radius of the combined mass of both black holes.  Middle: Inverse of the binary semi-major axis used as a measure of hardening rate. Plot elements are the same as on the left plot. Right: Eccentricity evolution of the SMBH binary.  Plot elements are the same as on the left plot.}
    \label{fig:bh_sep}
\end{figure*}

 In this section we present the orbital evolution of the SMBH binary and compare our results to the ones obtained in \citet{Khan2016}. We start the resimulation at $t=t\_0$ when the binary is hard and dynamical friction is no longer effective. At this point, the PN effects on the hardening are still insignificant, and the primary mechanism of energy loss for the binary is via the interactions with the stars from the loss cone.  On Figure ~\ref{fig:bh_sep} we present the orbital evolution of the SMBH binary compared to the original run  \citep[see Figure 3, left panel of][]{Khan2016}.  The hardening rate of the binary, defined as: 
\begin{equation}
s =\frac{d\big( 1/a \big)}{dt}   
,\end{equation} 
is approximately constant, which can be seen from the linear behaviour of $1/a$ in the middle panel of Figure ~\ref{fig:bh_sep}.  On the left panel of the same figure, we see the black hole separation as a function of time since $t\_{ini}$. The separation of the SMBH particles during our run is on the order of $\sim1000 R\_{sch}$. Up to resimulation time $t\_{0}$, we plot the data obtained with the $\varphi$-GPU code in \citet{Khan2016}. The period between  $t\_{0} < t < 32.35$ $\mathrm{Myr}$ corresponds to the time period of our resimulation of the system using $\varphi$-GRAPE-hybrid. 

From $t=32.35$ $\mathrm{Myr}$ and onwards, we plot the analytical estimates of the hardening driven by stellar interactions and GW emission, using well-known formulae in \citet{PetersMathews63} and a constant value of hardening rate $s = const$. The estimates were performed using standard Runge-Kutta fourth-order integration for three different values of hardening rates: $s= 4.5, 5.56, 6.4$ $\mathrm{Myr^{-1}\, pc^{\, 1}}$, where the middle value corresponds to the hardening rate that we measure at the end of the run (see Fig. \ref{fig:hists}, middle panel). From this, we estimate the SMBH merger timescale at $38.3\pm0.8$ $\mathrm{Myr}$ since $t\_{ini}$, or $16.8\pm0.8$ $\mathrm{Myr}$ since $t\_{PN}$. This value is a factor of two larger than the value reported in \citet{Khan2016}, who reported a merger timescale of under $10$ $\mathrm{Myr}$.
\par During our pilot runs we investigated different combinations of numerical parameter values to test how they affect the hardening rate. We found that the gravitational softening value and the minimum black hole integration time-step affected the hardening rate most significantly, resulting in the discrepancy between our merger timescale and the one in the original study.  The value of the merger time in \citet{Khan2016} might be underestimated primarily as a result of insufficient time resolution of the binary, determined by the minimum black hole integration time-step. This would have led to an overestimation of the PN terms in the equations of motion and a premature PN plunge. This is also visible in the eccentricity evolution, where we interpret the rise in eccentricity after $t = t\_{0}$ in the original data as evidence  of numerical artifacts. The decrease in the black hole minimum integration time-step reduced the effect of overestimation of the PN terms, and the hardening rate is now in agreement with  analytical formulae of GW emission \citep{PetersMathews63}, demonstrated by the dash-dotted lines in Figure ~\ref{fig:bh_sep}.
\par Additionally, we performed convergence tests in an effort to see if further decreasing the value of the softening from the original run affects the hardening rate. These tests showed a dependence of the hardening rate on the value of gravitational softening both using the $\varphi$-GRAPE-hybrid code, and the $\varphi$-GPU, which adopts an individual softening procedure and was used in the original study. We find that the hardening rate converged using both codes when the softening value for star--black hole interactions was set to $\epsilon \leq 2\times10^{-4}$ $\mathrm{pc}$, which is why we adopted this value for our run.  This suggests that the limited spatial resolution of the original run somewhat underestimated\footnotemark the hardening rate during close pericenter passages of stars. \footnotetext{It is important to note that, despite the underestimation of the hardening rate caused by the softening, the insufficient time resolution of the binary which overestimated the hardening rate was the dominant numerical effect that we believe resulted in the premature plunge driven by PN effects.}

\subsection{Merger remnant properties}
\label{sec:GMR}

\begin{figure}
\centering

        \includegraphics[width=\columnwidth]{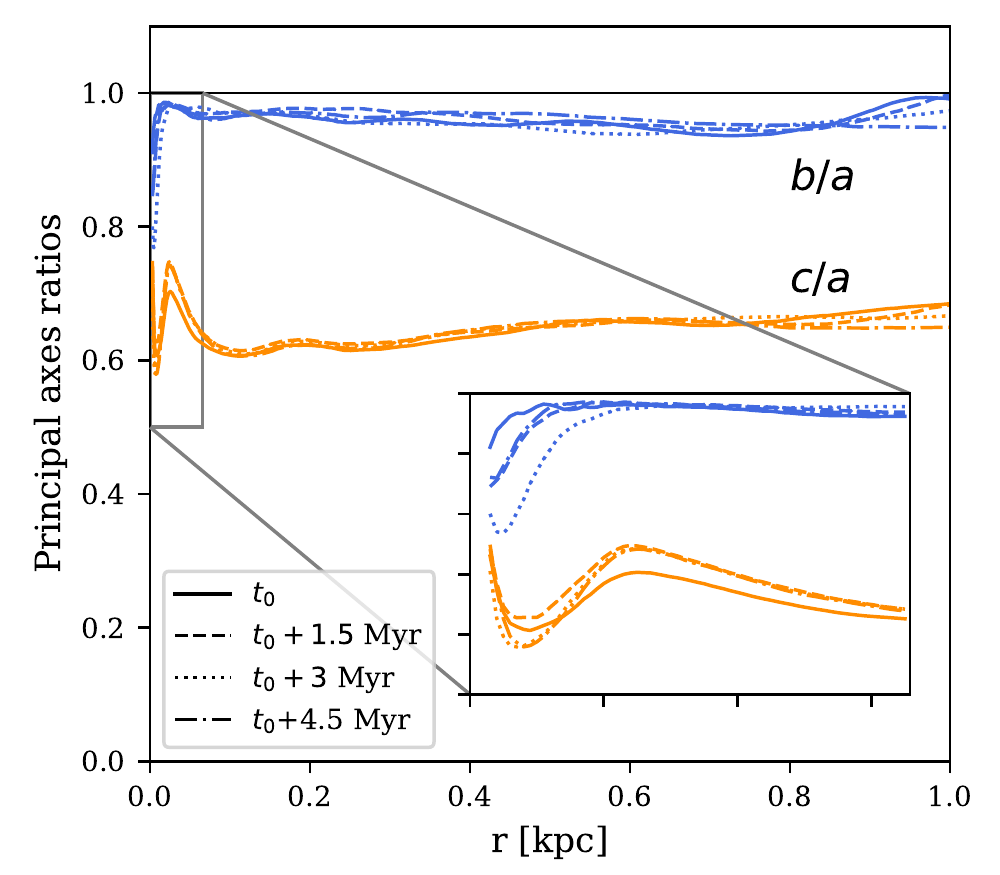}
    \caption{Ratios of the principal axes of the system up to $1$ $\mathrm{kpc}$ shown for different times. The figure in the bottom-right corner shows a zoomed-in region, ranging from $0.5$ to $1$ in the y-direction, and from $0$ to $5R\_{infl} = 66\, \mathrm{pc}$ in the x-direction, where $R\_{infl}$ is the SMBH binary influence radius.}  
    \label{fig:triax}
\end{figure}

\par In this section we present properties of the galaxy merger remnant. Our goal is to demonstrate the stationarity of the system in terms of its shape and orientation. This is done to make sure that the stellar parameters we analyze are not affected by changes in the galactic system and that the system does not measurably change its properties at the time of resimulation. In this way, our results would not be affected by the sudden change in the gravitational potential of the system as a result of the switching of the codes, and therefore the force calculation method. 

\par In the classical loss-cone regime inside of spherical nuclei, stars in the loss cone would interact with the SMBH binary on a crossing time timescale, after which the loss cone would be largely empty and the binary would stall \citep{Milosavljevic2003}. The refilling of the loss cone by two-body relaxation effects would happen on a relaxation timescale $t\_{relax}$ which may be larger than the Hubble time \citep{Berczik2005}. Therefore, collisionless refilling of the loss cone by stars on centrophilic orbits originating from triaxial or possibly axisymmetric orbits is necessary for efficient loss-cone refilling and a constant hardening rate. 

It is important to note that for the analysis of strong stellar interactions with the binary, we do not consider the first $1.8$ average crossing times ($1.55$ $\mathrm{Myr}$, which corresponds to 1 N-body time unit) and we only  use the last $3.1$ $\mathrm{Myr}$ of data for the analysis of energetic interactions. The reason for this is that, while the overall properties of the GMR do not change significantly at the moment we start the resimulation, owing to the potential switch it is possible that some stars would 
be artificially perturbed and put on centrophilic orbits. Therefore, during the first $1.55$ $\mathrm{Myr}$  we allow the system to relax and adjust to the new potential in order to ensure that all of the encounters we obtain are physical. 
\par In Fig. ~\ref{fig:triax}, we present a time evolution of the ratios of the principal axes of the system,  $b/a$ and $c/a,$ respectively, within a radius of $1$ $\mathrm{kpc}$, as well as the  inner-most region $5R\_{infl}$ ($\approx 66$ $\mathrm{pc}$, zoomed-in region). The principal axes were obtained from the eigenvalues of the following tensor\footnotemark: 
\footnotetext{The $I_{jk}$ tensor is sometimes referred to as the moment of inertia tensor, but this name is often reserved for the $I'_{jk}$ tensor, where $I'_{jk} = \mathrm{Tr}(I)\delta_{jk}- I_{jk}$  \citep{Binney1987}.} 

\begin{equation}
I_{jk} = \int \rho x_{j} x_{k}d^3x.
\end{equation}
We find the axis ratios in cumulative spheres in the range of $3$ $\mathrm{pc} < r < 5R\_{infl}$  for the main figure and in the range of $3$ $\mathrm{pc} < r < 1$ $\mathrm{kpc}$  for the zoomed-in figure.  The axes of the system were computed in the reference frame  that is comoving with the SMBH binary (described in Section \ref{sec:num}).  
Looking at the ratios of the medium and major axis, we can see that the galaxy merger remnant is slightly triaxial at all times ($b/a < 1$). It is precisely this slight triaxility that we expect to efficiently refill the loss cone, pointing to the fact that we should expect a roughly constant supply of stars in the SMBH binary vicinity, and therefore a constant hardening rate of the black holes as well. This is in agreement with the findings of \citet{Khan2016}, so we can conclude that there is no change in the shape of the system at the time of resimulation. 
Figure \ref{fig:triax} also shows us that the GMR is significantly flattened at all values of r with $
\epsilon \approx 0.4$. We calculate the flattening in the standard form $\epsilon = 1- c/a$, where $c/a$ is the ratio of the minor and major axis.

\par In order to investigate the orientation of the system, we first for each snapshot rotate the data so that the $z$-axis is aligned with the minor axis of the ellipsoid at $t\_0$ . We then look at the  orientation of the orbital angular momentum vector of the SMBH binary, as well as the cumulative angular momentum of the GMR system. To define the cumulative angular momentum of the system, we sum the contribution of all particles which are contained within a sphere of radius $r <5 R\_{infl}$, excluding the black holes. 
We find that during the time period of $3$ $\mathrm{Myr}$ used for the analysis of energetic interactions, the overall angular momentum vector of the system, and the smallest axis of the GMR ellipsoid match exceedingly well (the misalignment is always within $1 ^{\circ}$). This means that there is no misaligned rotation of the system, because the rotation of the GMR is aligned with the flattened shape, and that the three-body encounters are not scattered in a preferred direction because of this.

Additionally, the orientation of the SMBH binary orbit does not show significant changes during this period.  The orientation of the SMBH orbit is misaligned to the overall system by $\theta \approx 15^{\circ}$ throughout the run. It is expected that the SMBH binary orbital plane would realign according to the rotation axis of the system \citep{Rasskazov2017, Wang2014, Li2017} on a timescale of several hundred orbital periods \citep{Gualandris2012}. This realignment is caused by the angular momentum exchange and interaction with stars in a rotating cusp. Unlike these studies, we do not measure visible changes in the SMBH binary orbital plane. However, we argue that this is due to the already significant alignment with the system, and full realignment is not necessarily expected \citep{Gualandris2012}. Additionally,  the slight triaxiality and present rotation of our system might significantly reduce the alignment effect \citep{Cui2014}.

\par In order to properly characterize the orbits of stars from the loss cone, first we need to explore the properties of the overall system and understand the motion of stars within.  Because the GMR is only slightly triaxial, we assume axisymmetry for this part of the analysis. In the comoving reference frame, we switch to cylindrical coordinates given by $(R, \varphi, z)$  and explore properties in the $(R, z)$ (meridional) plane.

\par
\begin{figure}[h!]
        \includegraphics[width=\columnwidth]{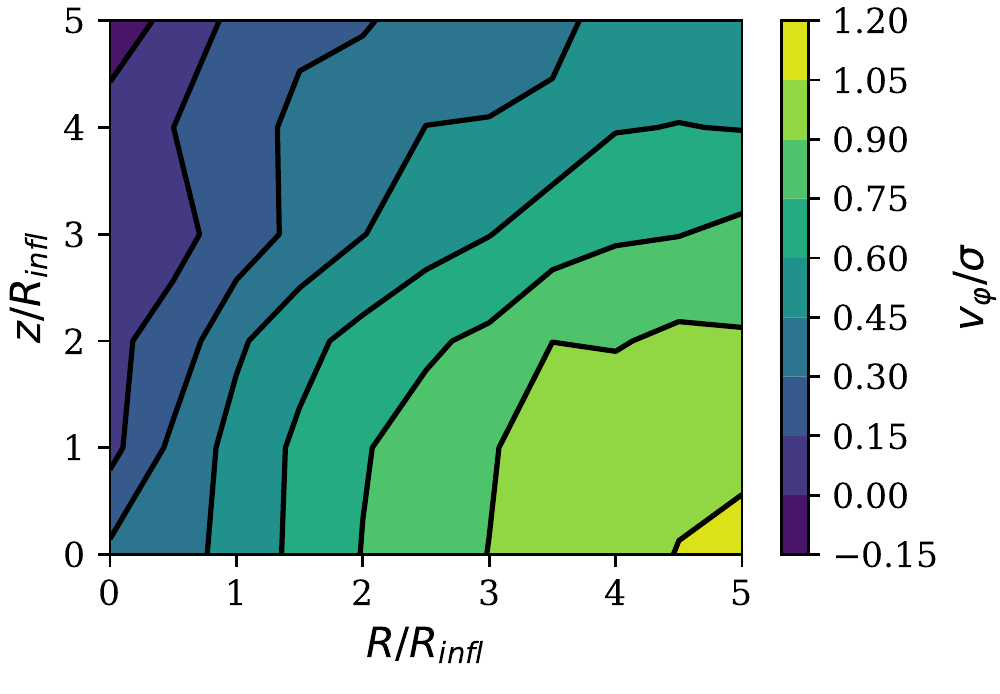}
    \caption{ Contour density plot of the tangential to random motion ratio $v_{\varphi}/\sigma$ in the meridional plane at the time of resimulation. 
    } 
    \label{fig:vphi}
\end{figure}

 In an oblate elliptical galaxy, the motions of the stars can be distinguished between a net streaming motion around the minor axis (rotation) and a random dispersion of velocities with respect to that streaming motion. The prevalence of one particular motion type  can offer important clues as to the orbital structure of the galaxy. With this in mind,  we start with a reference frame centered on the SMBH binary center of mass at time $t=t\_0$ (i.e., the comoving inertial reference frame described in Sect. \ref{sec:num}). We then rotate the frame of reference according to the three axes of the ellipsoid so that the minor axis is aligned with the overall angular momentum in direction $z$. Then, within $5R\_{infl}$  we divide the $(R,z)$ plane in a $10\times10$ grid of cylindric rings and find average values of kinematic properties of all the particles in each ring. In our reference frame, the streaming motion corresponds to the rotation around the z-axis, denoted by the tangential component of the velocity $V_{\varphi}$. The random motion is characterized by the velocity dispersion along each axis: $\sigma_{R}$, $\sigma_{\varphi}$, and $\sigma_{z}$. In  Fig. \ref{fig:vphi},  we can see the ratio of rotational to mean random motion in the meridional plane, where $\sigma = \sqrt{(\sigma_{R}^2 +\sigma_{\varphi}^2+\sigma_{z}^2)/3}$. We find that there is considerable rotation throughout the meridional plane, especially towards the equatorial plane. Our analysis shows us that after the galactic merger and throughout our run the GMR remnant retains a well-defined rotation axis that coincides with the minor axis of the remnant. This is also reflected in the motion of stars in the inner region, with a prevalence of ordered to random motion in the direction parallel to the equatorial plane.  

The strong rotation of the GMR undoubtedly plays a part in the significant flattening of the system that we measure, as shown in Fig. \ref{fig:triax}. In the past, rotation along the minor axis was considered the dominant contribution to the flattening of an axisymmetric system, however the discovery of giant elliptical galaxies not flattened by rotation showed that is not necessarily the case \citep{Binney2005, Bertola1977, Illingworth1977}. In reality, the flattening of an elliptic galaxy can be supported by rotation, significant velocity anisotropy, or a combination of both mechanisms \citep{Mo2010}. The prevalence of one or the other is often characterized by the anisotropy diagram, which relates the ratio of ordered and random motion $(V/\sigma)$ to the ellipticity of a galaxy. The tensor virial theorem in the formalism of \citet{Binney1987} gives us a relation between $(V/\sigma)$ and the flattening  ($\epsilon$) for an axisymmetric elliptical system that rotates about its symmetry axis (z-axis). This is accomplished with the $e'$ parameter \footnotemark which only depends on ellipticity: 
\footnotetext{This parameter should not be confused with orbital eccentricity $e$ which is used throughout the text. }     
\begin{equation}
e' = \sqrt{(1-(1-\epsilon)^2)}. 
\end{equation} 
Then, the  $(V/\sigma)$ relation takes the form \citep{Binney1987}: 
\begin{equation}
\label{eq: tvt}
(V/\sigma)^2 = 2(1-\delta)\Omega(e')-2, 
\end{equation}
where $\delta <1$ and $\Omega (e)$ in the oblate approximation takes the form of \citep{Cappellari2007}:
\begin{equation}
\Omega (e') = \frac{0.5(\arcsin{e' -e'\sqrt{1-{e'}^2}})}{e'\sqrt{1-{e'}^2}-(1-{e'}^2)\arcsin{e'}}. 
\end{equation}

\par The $\delta$ in Eq. \ref{eq: tvt} is a dimensionless global anisotropy parameter which quantifies the anisotropy between the symmetry axis and any direction orthogonal to it in an axisymmetric system \citep{Binney1987}.
\par
\begin{figure}[h!]

        \includegraphics[width=\columnwidth]{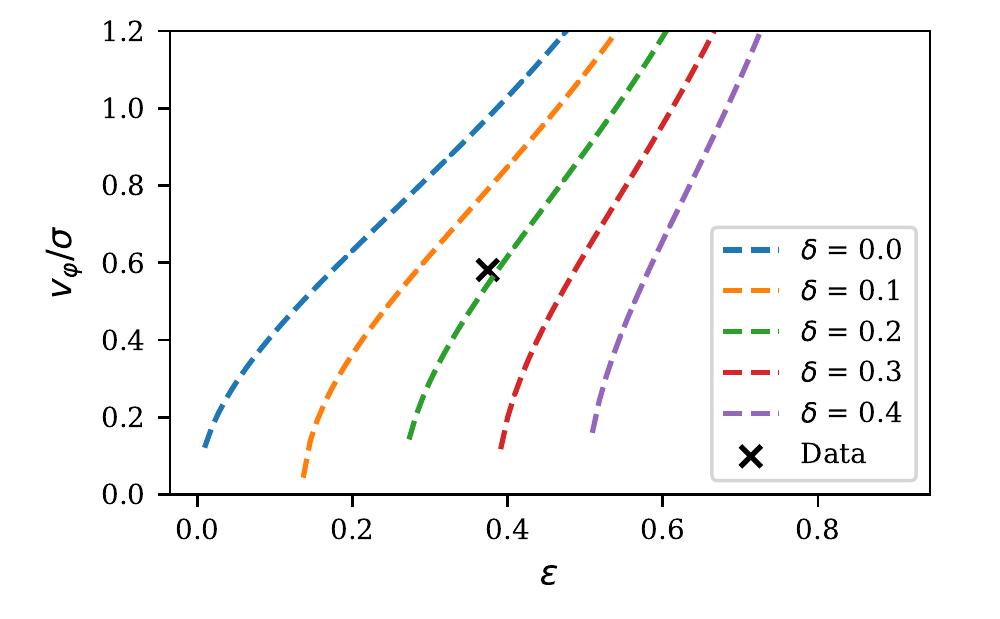}
    \caption{$(V/\sigma, \epsilon)$ relation from the tensor virial theorem. The lines correspond to different levels of anisotropy $\delta$. Our system is denoted by the black cross. }
    \label{fig:ani}
\end{figure}

 Relation \ref{eq: tvt} enables us to see where our system falls on the anisotropy diagram, and to what extent the observed flattening is supported by rotation around the minor axis (Fig. \ref{fig:ani}). We use the values of $\overline{v_{\varphi}}/\overline{\sigma}$ and $\epsilon$ at $R = 5R\_{infl}$ and $z=0$ as local analogs of the global values $V/\sigma$ and $\epsilon$. We can see that our system falls near the $\delta  = 0.2$ line in the plot, suggesting that while the flattening in our system is partially rotationally supported, rotation alone is not significant enough to account for the flattening. Therefore, there is also a non-negligible degree of anisotropy-supported flattening present. 

Additional insight into the flattening mechanisman of an  axisymmetric galaxy  can be gained with the $\gamma$ anisotropy parameter \citep{Thomas2009}, which we adopt from \citet{Cappellari2007}, along with the formula for $\delta$:

\begin{equation}
    \label{beta_gama}
 \delta = 1- \frac{2\sigma_{z}^2}{(\sigma_{R}^2+\sigma_{\varphi}^2)}  \quad
 \mathrm{and} \quad
 \gamma = 1- \frac{\sigma_{\varphi}^2}{\sigma_{R}^2}. 
\end{equation}

Using this formula for $\delta$, we obtain $\delta = 0.205$ at $z= 0$ for $R = 5R\_{infl}$, in very good agreement with the value estimated in Fig. \ref{fig:ani}. 

\begin{figure}[h]
  \centering
            \includegraphics[width=\columnwidth]{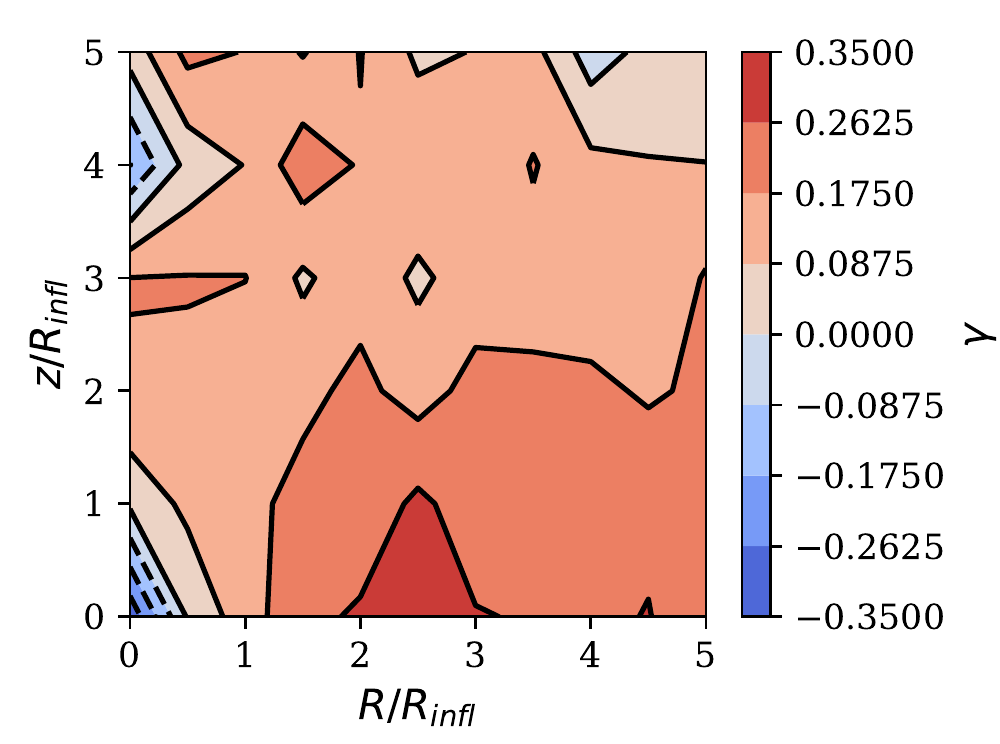}
      \caption{Contour density plot of the $\gamma$ anisotropy parameter, in the meridional plane at the time of resimulation.}
              
         \label{fig:kinematics}
\end{figure}

\par  Figure \ref{fig:kinematics} shows the anisotropy profile of the $\gamma$ parameter in the meridional plane at the time of resimulation $t\_0$.  We immediately notice that $\gamma >0$ throughout the meridional plane, suggesting a flattening supported in part by radial anisotropy. This is expected in merger remnants \citep{Thomas2007a} because of a large population of central box orbits which cause the centres of the merger remnants to become triaxial \citep{Thomas2009}.  Surrounding the SMBH binary, we see a prevalence of tangential orbits ($\gamma <0$), a clear sign that strongly radial orbits in the region have been scattered out and suggesting that the higher flattening up to $R\_{infl}$ (Figure \ref{fig:triax}, zoomed-in section) is supported by tangential anisotropy. The abundance of tangential orbits near the SMBH binary is in agreement with other studies that found the vicinity of the SMBH binary to have measurable tangential anisotropy \citep{Meiron2013,Meiron2010, Milosavljevic2001} with a large fraction of counter-rotating orbits with respect to the SMBH binary.

\subsection{Properties of centrophilic stars}
\label{sec:loss_cone}

\begin{figure*}[h!]

        \includegraphics[width=\textwidth]{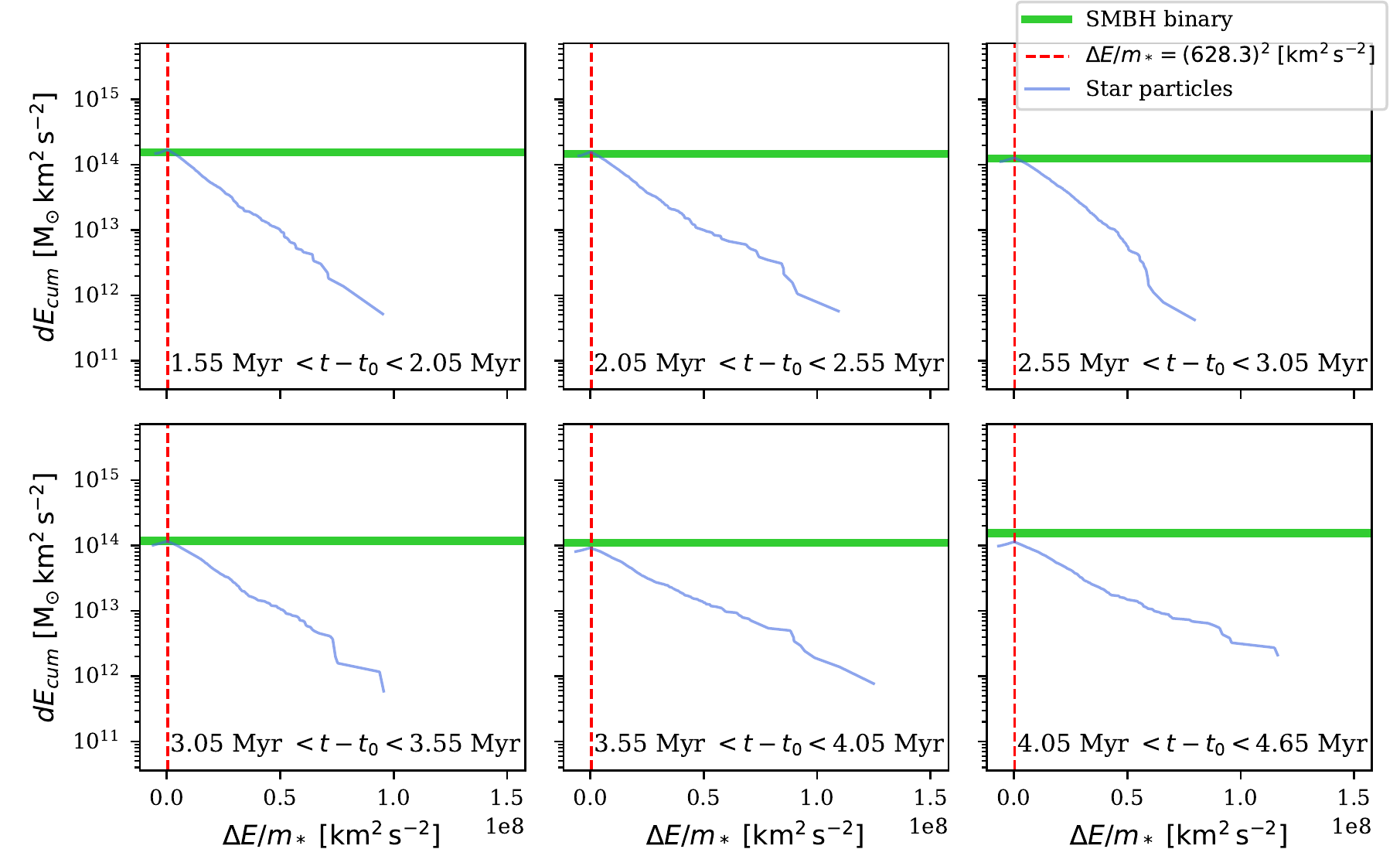}
    \caption{Energy balance plots showing the SMBH binary orbital energy changes compared to the cumulative energy changes of stellar particles for different times.   The blue lines represent the cumulative energy changes of all stars that come within $10a\_{bh}$ of the SMBH binary during the specified time period. The thick green line corresponds to the total SMBH binary orbital energy change for the same time period. The red dashed line corresponds to the cut-off value we use to define the high-energy tail.} 
    \label{fig:gap}
\end{figure*}

\par In this section we present our main findings and focus on exploring the parameters of the stars interacting strongly with the SMBH binary. During a strong interaction, the star experiences significant kinetic energy change, as it receives a strong velocity kick. Therefore, we only consider interactions where a star particle experiences large enough specific energy changes, which we refer to as the high-energy tail. We define the high-energy tail by finding stars with specific energy changes $\Delta E_{*}/m_{*} > (628.3)^2$ $\mathrm{km^2\,s}^{-2}$ during a single interaction.  This value is about $3\%$ of the total energy of the binary at resimulation time and is equal to 1 in N-body units; it was chosen because other effects produced from the overall evolution of the system could not reproduce such large energy changes. While relaxation effects could lead to changes in this value under certain conditions, in our simulation the vast majority of particles are treated with the SCF force, and the rest of the interactions are softened.   

\par To identify the interactions, we register passages within a sphere of $10a\_{bh}$ around the SMBH binary and monitor the energy changes.  In this way, we identify a total of 13383 strong interactions during a period of $3.1$ $\mathrm{Myr}$. We make sure that these strong interactions correspond to star particles and that dark matter particles do not contribute to the SMBH hardening in a measurable way. While most of the strongly interacting stars experience the potential in a direct way, we find that $20\%$ of these stars are \textit{halo} particles, and therefore the force they feel from other particles is calculated with the SCF method (except the force from the black holes which is always calculated directly). This fraction is in agreement with the percentage of halo particles we found in the inner region at the end of the run, namely  $\approx18\%$ , as described in section 2.2. We acknowledge that the presence of these \textit{halo} stars in the inner region is a consequence of the  assumption of spherical symmetry in our \textit{core--halo} criterion (Eq. \ref{eq:core_cond}), however we argue that this does not affect our results in a measurable way. The interactions with the black holes are unaffected by the \textit{core--halo} division and we find no statistical differences in the properties of strong interactions between \textit{core} and \textit{halo} stars.  We find that a star typically experiences multiple passages through the $10a\_{bh}$ sphere that result in small-angle and low-energy scatterings before finally experiencing a highly energetic interaction. With this in mind, we only consider the most energetic interaction per star for our analysis, unless explicitly stating otherwise. 
 \par Figure ~\ref{fig:gap} shows the cumulative energy changes of stellar particles during each passage through the $10a\_{bh}$ sphere around the binary, including nonenergetic interactions. The different plots correspond to different time intervals during the run, given in $0.5$ $\mathrm{Myr}$ increments. We compare all the cumulative energy changes of stars in the high-energy tail (to the right of the red dashed line) to the total SMBH binary orbital energy change during the same time period in order to make sure we are correctly identifying the stellar interactions that contribute to the binary hardening.  As we can see in the figure, the interactions we identify as the high-energy tail  match extremely well with the total orbital energy change of the SMBH binary at all times. Only in the final plot, just before we stop the run, do we notice a measurable gap in the energy balance, which is likely a result of insufficient accuracy in the black hole integration caused by the minimum integration time-step (see Section 4.3 for further discussion). We calculate the GW energy emission for these time intervals using the formulae from \citet{PetersMathews63} and we find that while measurable, the energy change from GW emission is several orders of magnitude smaller than the cumulative stellar hardening energy and the GW emission alone cannot account for the gap in the plot. Overall, the total SMBH binary orbital energy change over the entire period of $3.1$ $\mathrm{Myr}$ is $\Delta E\_{BH} = \Delta (-GM\_{BH1}M\_{BH2}/2a\_{bh}) = 4.07\times 10^{14}$ $\mathrm{M_{\odot}km^2s^{-2}}$, while the summed energy changes in stars in the high-energy tail is $E\_{cum} = \sum_{i=1}^{13383} \Delta E\_{max} = 3.701 \times 10^{14}$  $\mathrm{M_{\odot}{km^2s^{-2}}}$, where $\Delta E\_{max}$ corresponds to the energy extracted during their most energetic interaction. Therefore, we can conclude that our setup enables us to define the population of stars which contribute to the SMBH orbital energy changes with a very high degree of certainty.
 While additional effects could contribute to the overall binary energy evolution, such as for example gravitational torques  from overdensities in the mass distribution \citep{SouzaLima2020}, the good match between the energy extracted by interacting stars and the energy change of the binary show that stellar interactions can be considered the most important drivers of the binary hardening in the presented configuration, and other sources of energy change can be neglected.

\begin{figure}[h!]
        \includegraphics[width=\columnwidth]{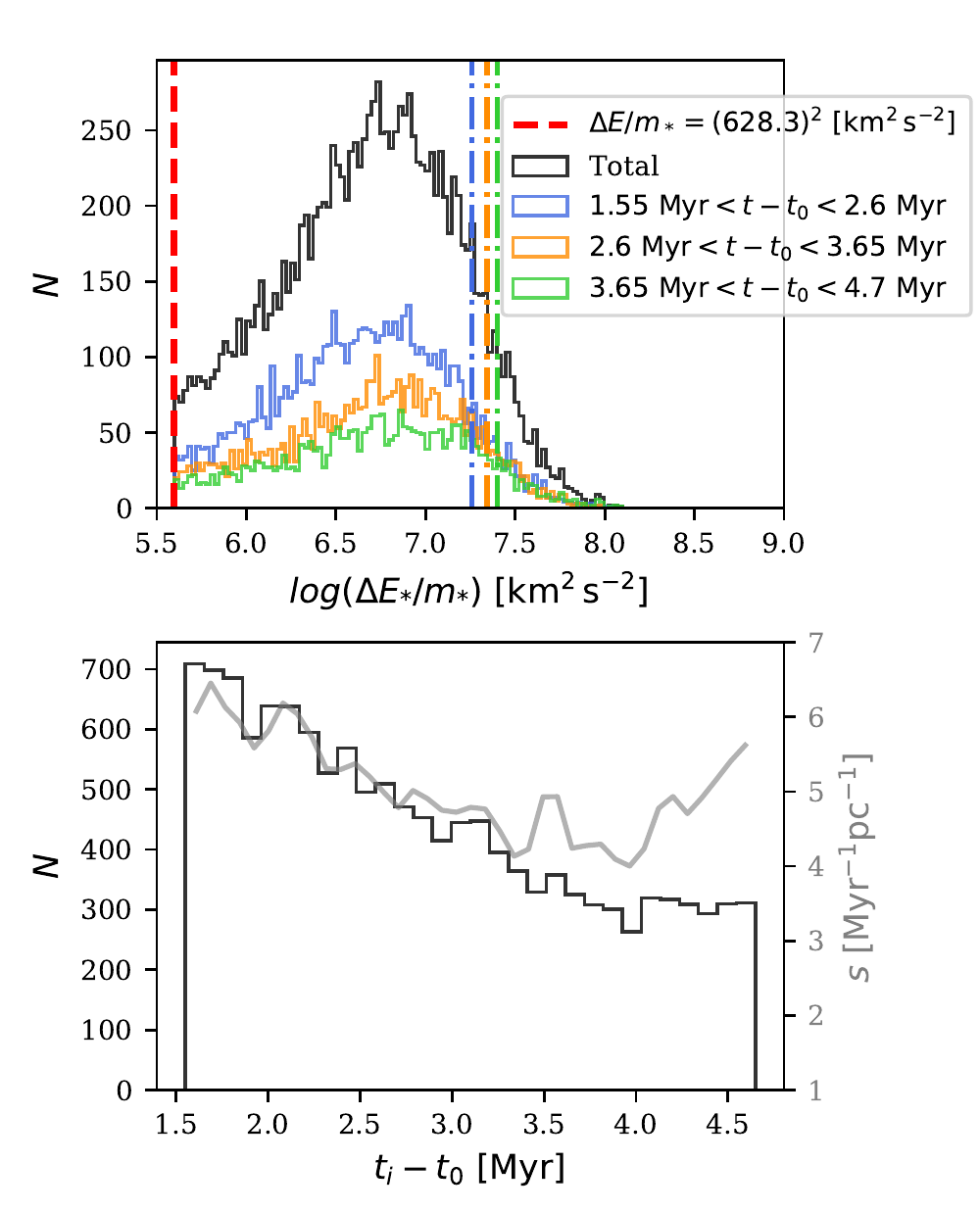}
    \caption{Top panel: In full lines, we present the distribution of specific energy changes in the interacting stars, where the colors correspond to different times throughout the run. The dash-dotted lines represent the expected typical specific energy changes of a star ejected by the binary at each time interval, given by Eq. \ref{eq:typ}. The red dashed line corresponds to the  cut-off value we use to define the high-energy tail.   Bottom panel: Time distribution of the most energetic interactions for each interacting star, given in $\mathrm{Myr}$ since $t\_0$. In gray, the hardening rate ($s$) is given as a measure of energy extracted from the binary by the interaction. We notice that while the number of encounters decreases, the hardening rate remains largely constant.  }
    \label{fig:hists}
\end{figure}

\par The distribution of the amount of energy the star particle extracts during a single encounter peaks at about $5\times10^{6}$  $\mathrm{km^2\,s}^{-2}$, which corresponds to $\Delta V = 2238$  $\mathrm{km\, s^{-1}}$ in terms of velocity kicks (Figure ~\ref{fig:hists}, top panel). This is slightly higher than the median velocity of the major SMBH, $\langle V\_{BH1}\rangle = 1519$ $\mathrm{km\, s^{-1}}$ and corresponds to $31\%$ of the median velocity of the minor SMBH $\langle V\_{BH2}\rangle = 5700$ $\mathrm{km\, s^{-1}}$. The typical specific energy changes of a star are  given in dash-dotted lines for different times in the same figure and are determined by the relation \citep{Merritt2013_b}:
\begin{equation}
  \label{eq:typ}  
   \Delta E\_{typ} = (M\_{BH2}/ M\_{BH})*\langle V\_{rel}\rangle,
\end{equation}
where  $M\_{BH2}$ is the mass of the least massive black hole,  $M\_{BH}$ is the total mass of the binary, and $\langle V\_{rel}\rangle$ is the median relative velocity of the binary during that time interval. 
The bottom panel shows that the overall number of interactions decreases with time. However, as the binary semi-major axis shrinks with time, the energy per encounter increases, leading to a roughly constant hardening rate, as demonstrated by the gray line in the same plot.

\begin{figure}[h!]
    \includegraphics[width=\columnwidth]{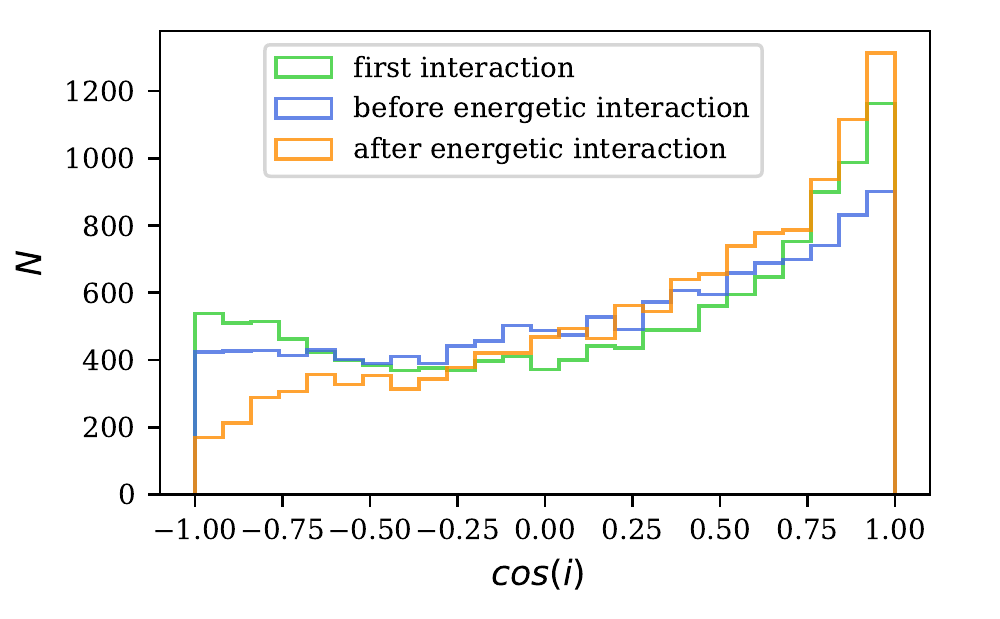}

        \includegraphics[width=\columnwidth]{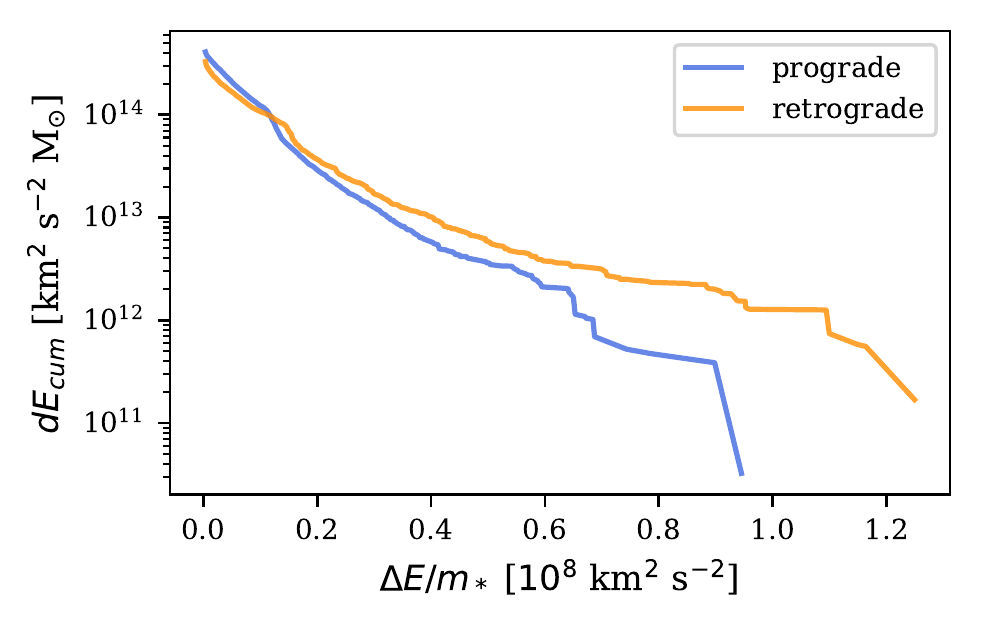}
        \includegraphics[width=\columnwidth]{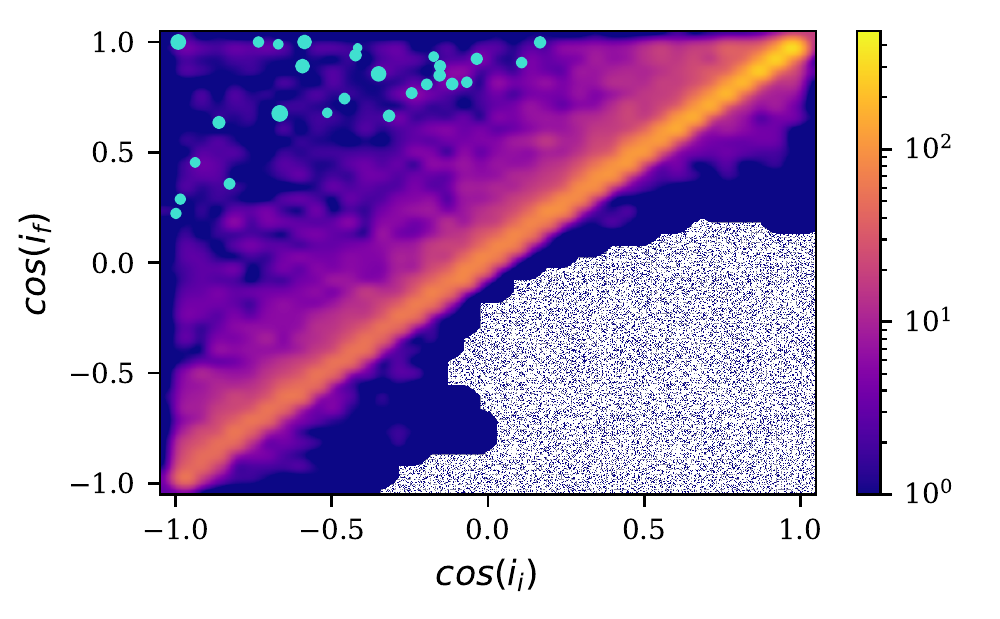}
    \caption{Top: Distributions of orbital inclination of the encounters with respect to the black hole binary, at first passage (green) as well as before and after the energetic interaction (blue and orange lines, respectively). Middle: Cumulatively summed maximum energy changes in prograde (blue) and retrograde (orange) orbits as a function of specific energy change. Bottom: Two-dimensional histogram of the initial (x-axis) and final inclination (y-axis). The green points correspond to the 27 most energetic encounters (with $\Delta E/m_* > 7.9\times10^7$ $\mathrm{[km^2 s^{-2}]}$). The size of points is correlated with the total energy extracted.}
    \label{fig:incl_groups}
\end{figure}

\par We can estimate the incoming inclination of  interacting stars entering the $10a\_{bh}$ sphere with respect to the SMBH binary orbital plane from the orientation of their respective angular momentum vectors (top plot on Fig. \ref{fig:incl_groups}). We find that there is a preference for prograde rotating stars with respect to the SMBH binary ($\cos(i) \geq 0$) compared to the population of stars with retrograde rotation ($\cos(i) < 0$). While there are no significant changes in orientation just before the first and final interaction (green and blue lines in top panel of Fig. \ref{fig:incl_groups}), we do notice that there are a number of retrograde stars that experience an angular momentum  sign-flip change during their energetic interaction, resulting in them becoming prograde (orange line in the same plot).

\par However, let us focus our attention on the middle part of the same figure, which shows the cumulative energy change of prograde ($cos(i\_i) \geq 0 $) and retrograde ($cos(i\_i)<0$) interactions as a function of specific energy change. While the prograde interactions are much larger in number, they account for only slightly more in the cumulative energy change than the retrograde group. This is because of the fact that the retrograde interactions dominate the high-energy tail of the distribution, as evidenced by the right side of the plot. Therefore, retrograde interactions, while lower in number, account for the highest energy changes. This is evident in the bottom panel of Fig. \ref{fig:incl_groups}, where the 27 most energetic encounters are given as green points. Most interactions do not experience any visible changes in their inclination and therefore fall on the diagonal on the plot. The most energetic interactions are almost all initially retrograde, and experience the largest inclination changes, orienting according to the SMBH binary orbital plane. This is due to the fact that the energy change of a star during the encounter is proportional to its change in angular momentum parallel to the SMBH binary orientation \citep[Eq. 63 in][]{Rasskazov2017}. 

\begin{figure}
\centering
    \includegraphics{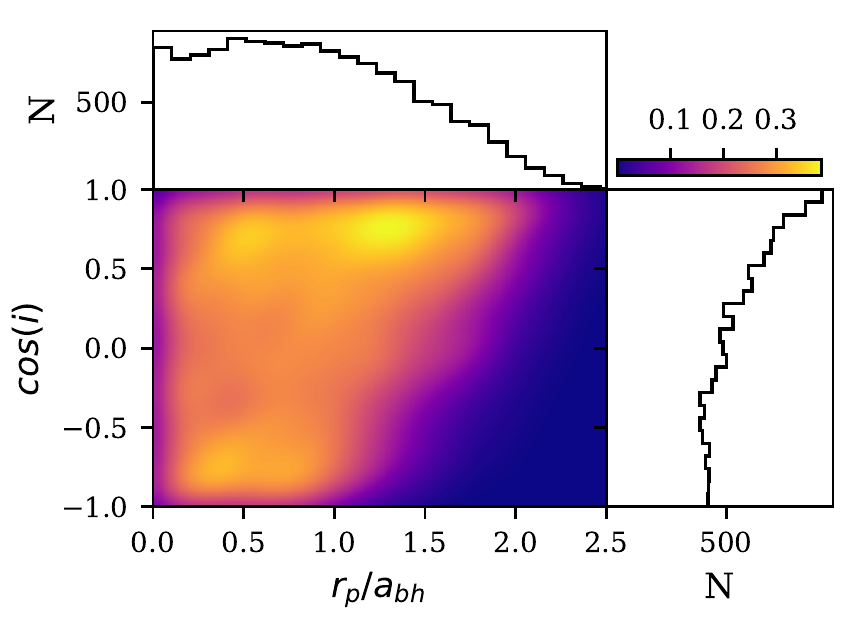}
    \caption{Density distribution of the inclination of the stellar orbit passage with respect to the SMBH binary orbital plane (y-axis) and the Keplerian pericenter (x-axis) at the time of the energetic interaction, normalized to the binary semi-major axis value.  The distribution was smoothed using a Gaussian kernel density estimation.}
    \label{fig:2dhist_rp}
\end{figure}

\par The pericenter distance of a stellar orbit is a significant parameter that facilitates strong energy exchange. In order to experience a sufficiently strong energetic interaction it is expected that the star should come within several $a\_{bh}$ of the SMBH binary. We can confirm this by looking at Fig. \ref{fig:2dhist_rp}, where we see that for all of the energetic interactions the stars come within $3a\_{bh}$ of the binary. Generally, the closer the encounter is to the binary, the more energetic it will be. It is then no surprise to see that the retrograde interactions have much lower pericenter values. In fact, we find that unlike the prograde group, almost all of the retrograde interactions cross the binary orbit $r\_{p}< a\_{bh}$. 
Additionally, we obtain no highly energetic interactions with $r\_p > 3 a\_{bh}$. Therefore, we find this value to be a hard boundary for the pericenter of the energetic interactions in our run. However, we note that the pericenter shown here is calculated from a Keplerian approximation which ignores the binarity of the SMBH system, and is meant only as a proxy to estimate the shape and size of the stellar orbit from a distance of $10a\_{bh}$ from the binary. The actual three-body interaction is far more complex and depends on the orbital phase of the binary.

\begin{figure}[h!]

        \includegraphics[width=\columnwidth]{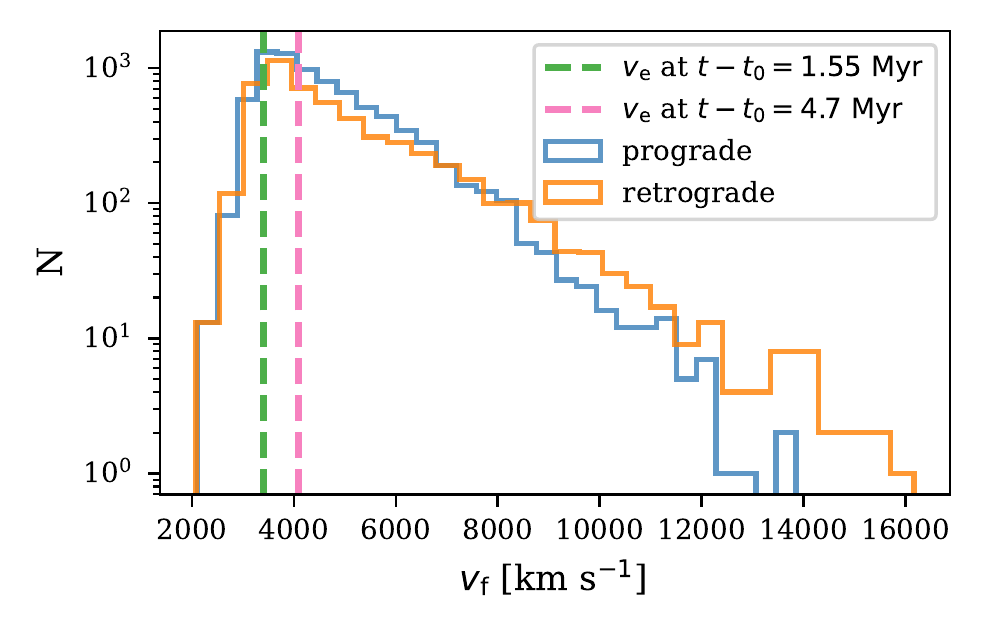}

    \caption{Ejection velocity distribution of the interacting stars with respect to the SMBH binary. Blue and orange lines correspond to stars on prograde and retrograde orbits with respect to the binary orbital plane, respectively. The dashed lines represent the escape velocity from the overall galactic system at a radius of $10a\_{bh}$ from the binary center of mass, at the start and at the end of the run. }
    \label{fig:hist_vf}
\end{figure}

Similarly, we can look at the ejection velocity distributions of the prograde and retrograde interactions (Figure \ref{fig:hist_vf}). The ejection velocities for both populations are peaked just beyond the escape velocity (with respect to the galactic system) at $r=10a\_{bh}$ from the binary, which ranges   from $V\_e = 3410$ $\mathrm{km\, s^{-1}}$ at the beginning of the analysis ($t-t\_{0} = 1.55$ Myr) to $V\_e = 4093$ $\mathrm{km\, s^{-1}}$ at the end of the run.  We find that the prograde distribution is narrower and that the retrograde distribution extends up to $16,000$ $\mathrm{km\, s^{-1}}$. This behavior is in agreement with our results so far and the velocity distribution presented in \citet{ArcaSedda2019}, although we obtain significantly higher velocities than these latter authors, likely as a result of our simulation having more massive SMBH particles with lower separation. 

\begin{figure}
\centering
        
        \includegraphics[width =\columnwidth]{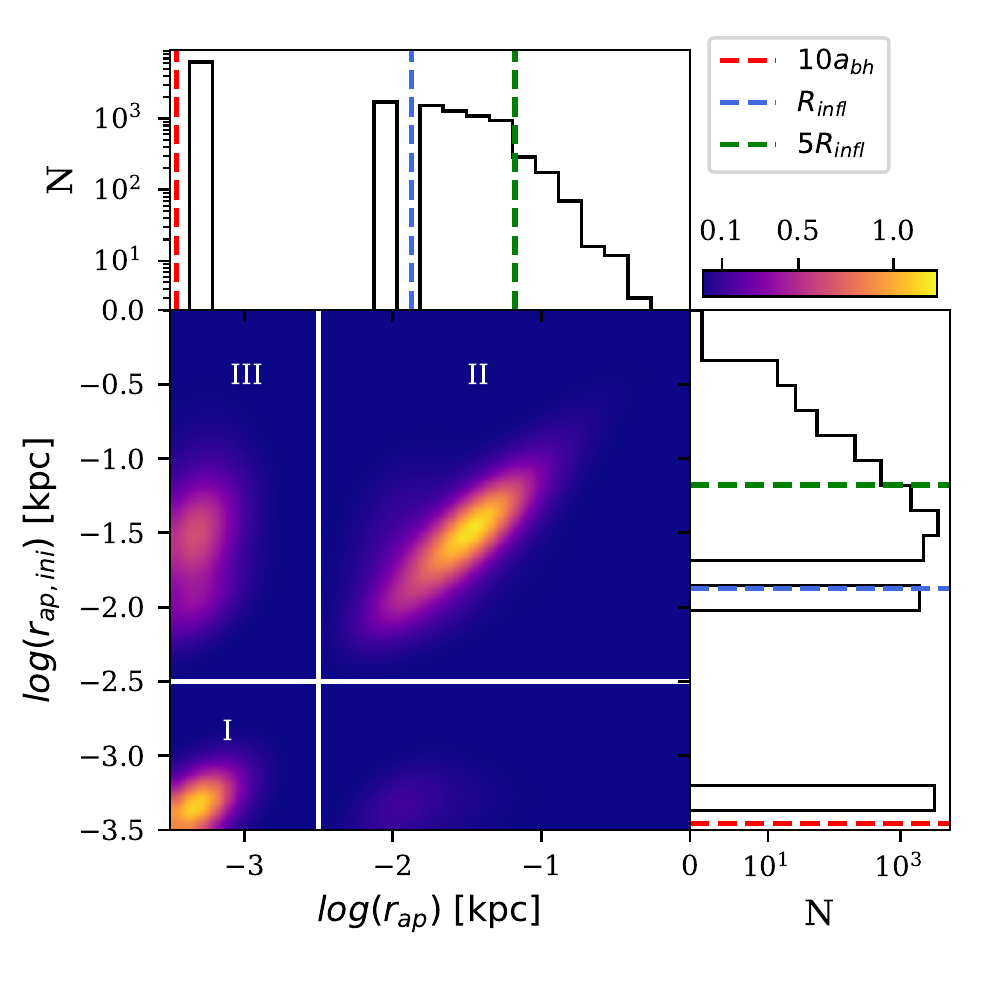}

        \includegraphics[width =\columnwidth]{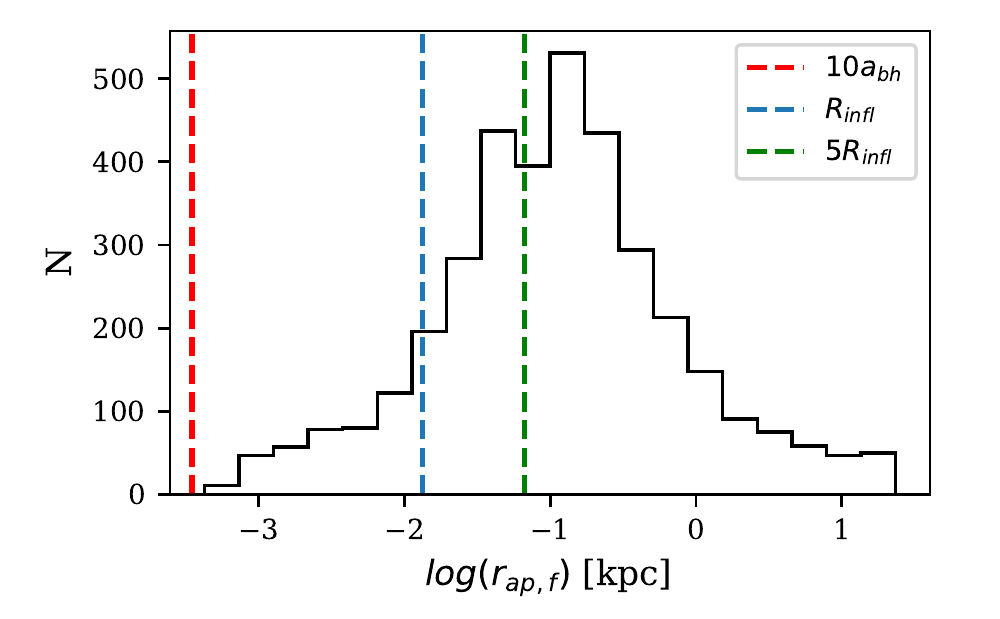}
    \caption{Top: Apocenter distribution of stellar orbits just before the first interaction (y-axis)  and just before the most energetic interaction (x-axis), in log-spaced bins.  The distribution was smoothed using a Gaussian kernel density estimation. Bottom: Apocenter distribution of stars still bound to the galactic system after the most energetic interaction with the SMBH binary.}
    \label{fig:rap_log}
\end{figure}

Loss cone orbits are defined in terms of their position in phase space rather than in physical space. This is because even stars well outside the SMBH binary region of influence can be part of the loss cone if their angular momentum is low enough. However, to what radius in physical distance a potential loss cone orbit can extend remains unclear.  In order to investigate this, we estimate the apocenter of the stellar orbits by assuming that energy is conserved along the orbit. This is a justified approximation because the orbits are highly eccentric and the kinetic energy at apocenter is negligible.  We estimate the apocenter by finding the maximum radius a particle with that energy can have in the potential of the system, assuming spherical symmetry for the potential.  In this way, our approximation of the apocenter is only dependant on the energy of the stellar particle, unlike our pericenter approximation which takes into account both the energy and angular momentum of the orbit.  We present the estimates of the apocenters of the star at the time of the first  (not necessarily energetic) interaction (y-axis) and at the time just before the most energetic interaction (x-axis) in the top panel of  Fig. \ref{fig:rap_log}. In the 2D histogram, we notice three distinct populations, which are also evident in the top and right histograms. The two populations on the diagonal line (hereafter Populations I and II, from left to right) consist of encounters where there is no visible change in the apocenter before the energetic interaction, either as a result of  multiple low-energy interactions that do not change the orbital parameters significantly (Pop. I) or because they consist of single energetic interactions (Pop. II). The gap between these populations shows us that the loss cone is essentially empty within $R\_{infl}$, except for the population very close to the SMBH binary. The population outside of the diagonal (hereafter Population III) initially comes from within $R\_{infl} < r\_{ap} < 5R\_{infl}$ and changes its orbital properties in a number of weak three-body encounters, decreasing the energy, and as a result also the apocenter, until the pericenter falls below $3a\_{bh}$ and it experiences an energetic interaction.  While there are some outliers, we can see that almost all of the interactions come from a region within $5R\_{infl}$, or $66$ $\mathrm{pc}$ in physical units, and after this point the distribution drops quite sharply.

After the interaction, the stars get a kick in their orbital velocity and some gain enough energy to leave the system as a HVS. We find that a total of $72\%$ of interacting stars have a positive total energy after the interaction $E\_{tot}>0$, and thus can be ejected from the system completely. The remaining $28\%$, or 3649 stars ($\approx 1000$ in each population),  still gain a significant velocity kick, and we estimate their new apocenters to investigate the nature of their orbits. In the bottom
panel of  Fig. \ref{fig:rap_log}, we see that the maximum of the distribution lies just beyond $5R\_{infl}$. As the vast majority of our interactions  are contained within the region of $5R\_{infl}$ at initial apocenter, we do not expect this population to be able to return and interact strongly with the binary another time. However, the population contained within $r\_{ap} < 5R\_{infl}$ still has that possibility, because the radial orbital timescale at $R=5R\_{infl}$ is $\sim5\times10^4$ $\mathrm{yr}$.   Some stars, while still bound, are put on exceedingly extended orbits, with apocenters going up to $23$ $ \mathrm{kpc}$ in the high $r$ tail of the distribution.

\begin{table*}
\centering
\caption{General properties of stellar encounters of  Populations I-III.}
\begin{tabular}{c | c | c | c | c  }
\hline \hline
  &$ N$ &  $ \langle \Delta E/m  \rangle$  & $ \langle dt  \rangle$   & $ dE\_{cum}$    \\
    & &   $[\mathrm{km^2\, s^{-2}}]$ & $[\mathrm{yr}]$ &  $[M_{\odot}\,\mathrm{km^2\, s^{-2}}$]  \\ 
 \hline
   \textit{Total} & 13383  & $ 5.11\times 10^6 $ & $ 153.73 $  & $7.36\times 10^{14}$ \\
  \hline
   \textit{Population 1} & 2816  & $ 4.51\times 10^6 $ & $ 177.57 $ & $1.77 \times 10^{14}$ \\
    \hline
   \textit{Population 2} & 6680  & $ 4.79\times 10^6 $ & $ 138.95 $ & $3.33 \times 10^{14}$  \\
     \hline 
        \textit{Population 3} & 3465 &  $ 5.93\times 10^6 $ & $ 166.94 $ & $1.95 \times 10^{14}$ \\
    \hline \\

\end{tabular}
\tablefoot{The columns represent, from left to right: number of stars in the population, median specific energy change per encounter, median duration of most energetic encounter,  and cumulative energy change of the population. The values of the three rightmost quantities are calculated from only the most energetic interaction of each star.}
\label{tab:tab_mean}
\end{table*}

\par In Table \ref{tab:tab_mean} we present the general characteristics of stellar encounters in the apocenter populations I-III. The 422 stars which do not belong to any population from Figure \ref{fig:rap_log} (top panel) were not considered here because of their low number statistics and negligible contribution to the binary  hardening. In the table we notice that Pop. II is the most populous, with 6680 stars, and also has the shortest median duration of the most energetic encounter. This population contributes most significantly to the hardening, with $45\%$ of the overall energy exchange. Population III, while smaller in number, has the highest median specific  energy change, suggesting more energetic interactions. On the other hand, Population I is the least populous, has the lowest median specific energy change per encounter, and therefore contributes the least to the overall binary hardening. 

\par
\begin{figure*}
\centering
        
        \includegraphics[width =\textwidth]{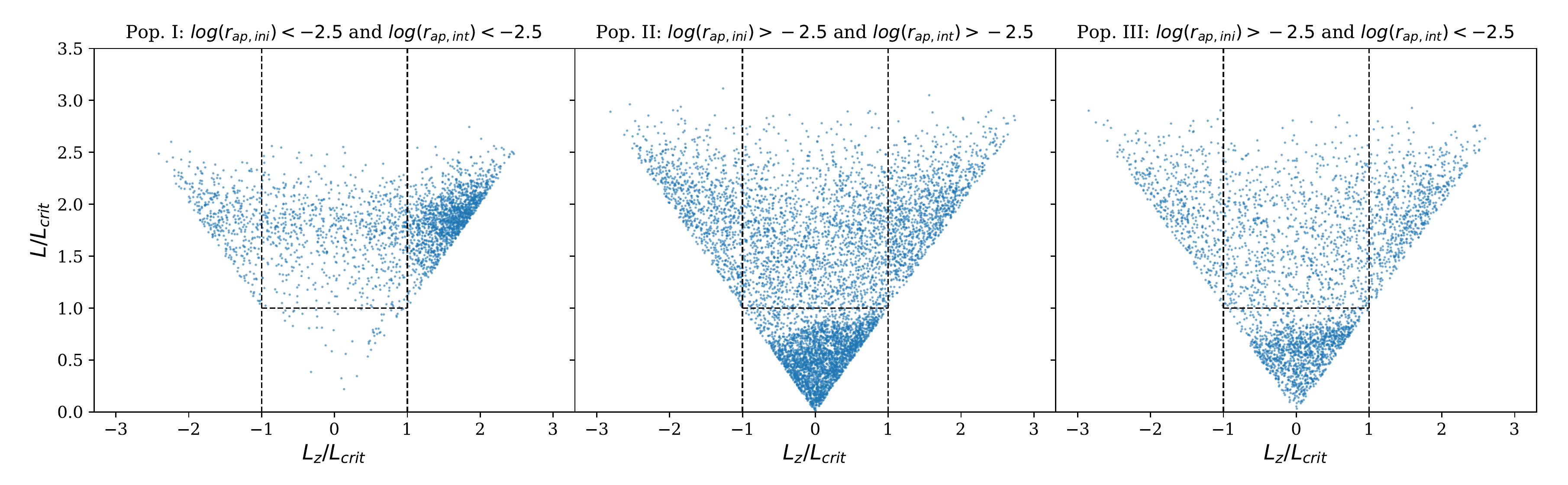}

        \includegraphics[width =\textwidth]{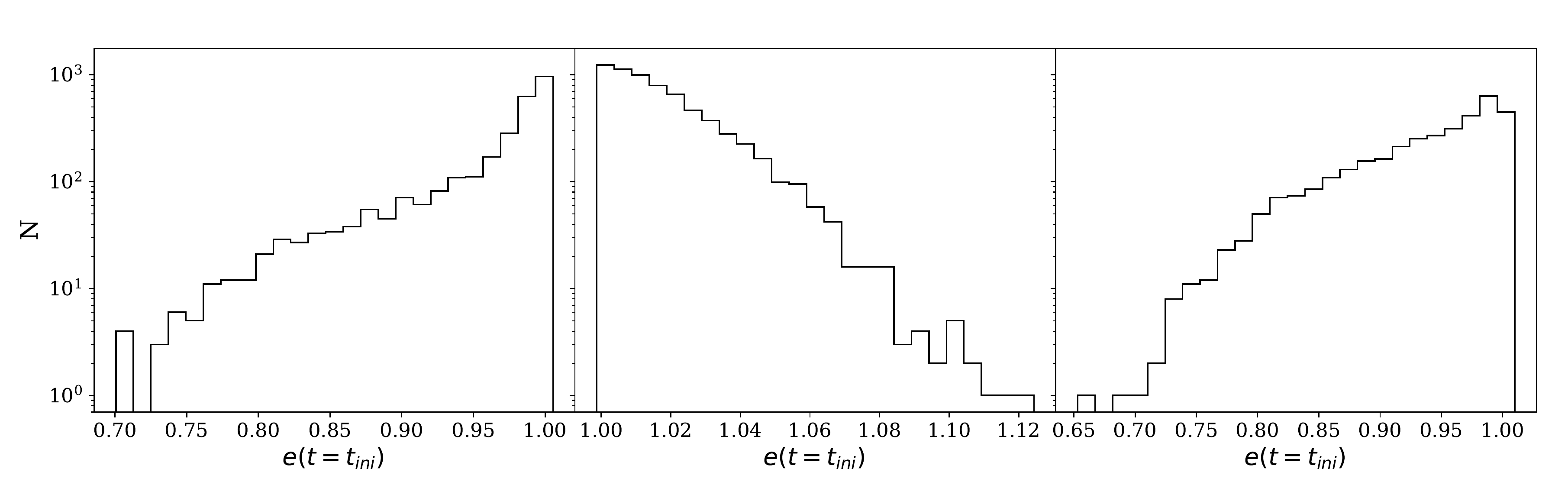}

        \includegraphics[width =\textwidth]{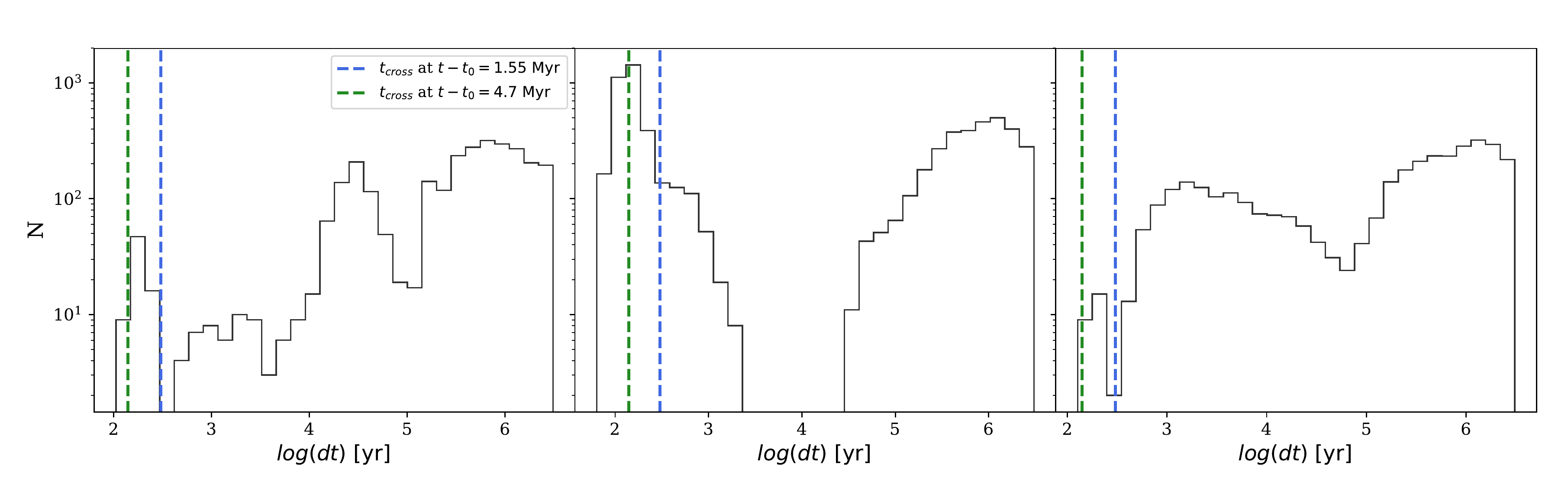}

    \caption{Top: Phase space properties of Populations I-III at the time of their first interaction with the SMBH binary. The bottom rectangular region bounded by dashed lines of $L = L\_{crit}$ corresponds to the classic loss cone, when $\eta = 3$. Middle: Eccentricity distributions in the Keplerian star-SMBH binary approximation at the time of first interaction.  Bottom panel: Total duration of all of the interactions per particle (including non-energetic interactions), given in years and calculated as the amount of time between the first registered entrance and final exit from the sphere of radius $r=10a\_{bh}$ around the black hole binary. The dashed lines correspond to the central crossing time of a star on a parabolic orbit for different times.} 
    \label{fig:lini_box}
\end{figure*}

 The populations can be characterized by taking a look at their phase space distributions. Namely, the classical loss cone of an SMBH binary is typically characterized as the region populated by low-angular-momentum stars, which follow the condition \citep{Yu2002}:
\begin{equation}
\label{eq:lc}
L \leq L\_{crit} = \sqrt{(\eta2GM\_{BH}a\_{BH}},  
\end{equation}
where $\eta$ is a dimensionless factor on the order of unity and assuming spherical geometry. 
 However, in nonspherical nuclei, there is an extended region in phase space consisting of stars which, while not originally in the loss cone, may be driven into the classical loss cone by nonspherical torques. This region has previously been referred to as the loss region \citep{Vasiliev_2014}.  We  plot the distribution of the three apocenter populations (Pop. I-III)  in the $(L, L_z)$ plane in Fig. \ref{fig:lini_box}, at the time of their first passage, taking the value $\eta = 3$ (which is agrees with our pericenter investigations). The $z$ in $L_z$  corresponds to the minor axis of the GMR.  We immediately notice that in all three cases, there is a large region of parameter space populated by stellar orbits, which are not yet in the classical loss cone at the time of their first passage (denoted by the bottom dashed square at $L/L\_{crit} <1$ and $-1<L_z/L\_{crit}<1$). These correspond to the stars which interact with the binary several times before their energetic interaction and consequent ejection. We find that at the time of their most energetic interaction, all of them satisfy condition $\ref{eq:lc}$, with $\eta = 3$. 
 
 \par The first population, situated close to the SMBH binary, shows significant positive rotation around the z-axis, corotating with the binary as well as the system during their first encounter.  Interestingly, we see no visible difference in the phase space properties of the Pops. II and III, both of which have initial apocenters  $\log(r\_{ap,ini}) > -2.5$ $\mathrm{[kpc]}$. Both populations are skewed towards positive $L_{z}$, suggesting a preference for corotating stars, in agreement with Fig. \ref{fig:incl_groups}. However, unlike the first population, most of the stars have sufficiently low angular momentum to be considered proper loss cone stars. It is therefore puzzling that only the second population is promptly ejected at this time, while the third one is captured by the binary. The reason for this difference lies in their energies with respect to the SMBH binary. We can see in the middle panels of Fig. \ref{fig:lini_box} that the second population is almost completely unbound to the binary, and therefore represents a population of fast, parabolic, and hyperbolic encounters. On the other hand, Pop. III is captured by the binary during its first passage, most likely by chance due to the incoming orbital phase with respect to the binary,  and the stars are put on eccentric orbits before their ejection, despite having the same phase space properties as Pop. II. 
 \par  The bottom panels of Fig. \ref{fig:lini_box} can provide insight into which stars experience several interactions and which interact only once. For this purpose, we show  histograms of the total interaction time per stellar particle, including all nonenergetic interactions with the binary. Namely, the duration of the majority of single interactions would be quick slingshots with small pericenters, which should correspond to crossing times of stars on parabolic orbits, denoted by dashed lines on the figure. We see that Pops. I and III have typically long interaction times, ranging up to the entire duration of the run ($\sim 10^6$ $\mathrm{Myr}$) and suggesting that they have many multiple passages before experiencing a strong interaction. Population II on the other hand, shows the strongest peak at shortest interaction times,  making this population characterized by single interactions: although, there is also a second peak at $dt \sim 1$ $\mathrm{Myr}$. Unlike the other two populations, there is a prominent gap between the two peaks, which might signify that this group consists of stars that have already experienced a strong interaction, and have returned to interact once more, after a few crossing times of the system ($t\_{cross} \sim 1$ $\mathrm{Myr}$).  
 \par
\begin{figure*}
\centering
        
        \includegraphics[width =\textwidth]{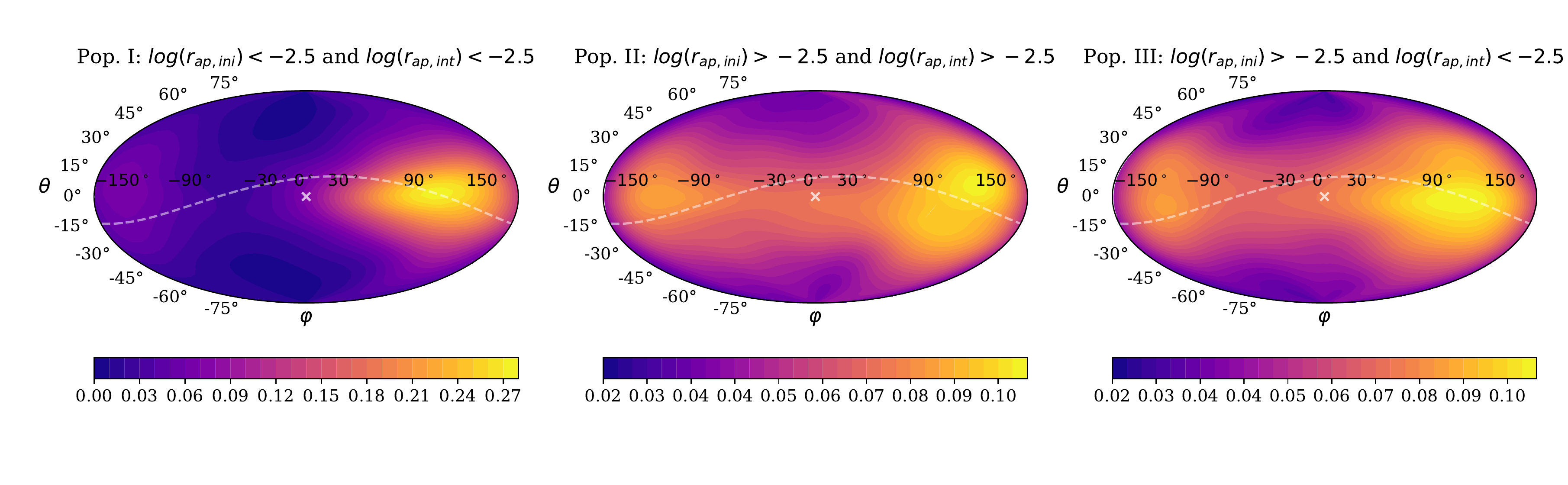}

        \includegraphics[width =\textwidth]{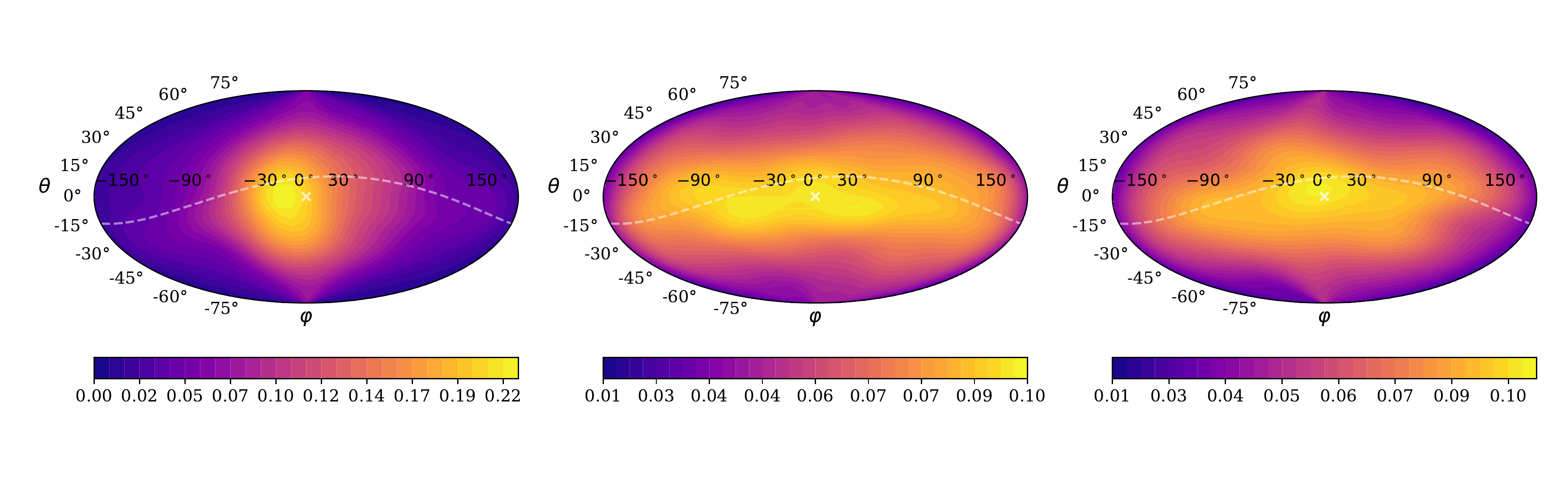}
        
    \caption{Top: Angular ejection distribution of Populations I-III at the time of initial interaction with the binary. The figures are centered on the SMBH binary center of mass, with the gray line representing the projection of the SMBH binary orbit. The white cross  corresponds to the direction of the origin of our comoving reference frame, where the potential is evaluated.  The $\theta = 0$ plane corresponds to the equatorial plane of the system. Bottom: Angular ejection distribution after the most energetic interaction. Other elements in the figures are the same as in the top figures.} 
    \label{fig:ejec_angle}
\end{figure*}

 Figure \ref{fig:ejec_angle}  presents the spatial angular distributions of stars in an equal area projection,  of incoming stars at initial passage (top  panels), and after the most energetic interaction (bottom panels), centered on the SMBH binary center of mass. We immediately notice that Pop. I is highly anisotropic in the azimuthal angle, both before and after interaction. Our investigations show that this is the result of the Brownian motion of the black holes around the centre of our comoving reference frame. Namely, we observe that the black holes experience a Brownian motion  around the center of the system (in the comoving reference frame) with an amplitude of  $\approx 1$ pc. Unlike \citet{Khan2020}, we do not observe Brownian motion in a rotational fashion despite the rotation of our system. Instead, the Brownian motion we measure is in the form of a random walk. The stars in Pop. I belong to an inner, rotating stellar cusp. Stellar hardening is known to erode the central stellar cusp, leaving lower densities in the very center \citep{Khan2012a}. As a result,  as the black holes move around the center in a random walk,  they preferentially disrupt and capture the population that surrounds this inner region, which will be in the opposite direction from the origin when viewed from the perspective of the black holes (Fig. \ref{fig:ejec_angle}, top-left panel). We therefore conclude that the presence of this population in the loss cone is the combined result of very high central densities of our system and the Brownian motion of the black holes.

 In contrast, we find that Pops. II and III are more isotropically distributed. These are the stars that come from outer regions on centrophilic orbits, and therefore the effect of random motion of the black holes does not play a role, like it does in the case of Pop. I. We notice that the third and especially second population show a prevalence of ejections along the SMBH binary orbital plane. This is in agreement with previous studies which find that HVSs are primarily ejected near the SMBH binary orbital plane when their SMBH orbital plane is aligned with the rotational place of the system \citep[e.g.,][]{Sesana2006, Wang2014, Zhong2014, Lezhnin2019,  Rasskazov2019HVS}.  Our findings show that not all of our three populations are uniformly distributed in the azimuthal direction.  While there is some disagreement in the literature over the existence of a preferential ejection direction in eccentric binaries \citep{Rasskazov2019HVS}, in the case of a circular SMBH binary like ours, previous studies report a uniform distribution of ejected stars in the orbital plane \citep{Rasskazov2019HVS, Lezhnin2019, Sesana2006}. We argue that the motion of black holes around the center can result in anisotropic ejections for the central population of stars (Pop. I), while stars originating from outside of the SMBH binary influence sphere are uniformly distributed (Pop. II and III).

\par From the apocenter distribution in Fig. \ref{fig:rap_log}, we conclude that Populations II and III correspond to stars on centrophilic orbits that are responsible for repopulating the loss cone. Depending on the shape and degree of nonsphericity of the nuclei (e.g., axisymmetric or triaxial), different orbital families may fulfil this role \citep{Merritt2013}.  We can distinguish between the different orbital families  if we look at their angular momentum at a time significantly before the interaction. Namely, because of the symmetries of different geometries, we can distinguish between \textit{spherical}, \textit{axisymmetric,} and \textit{triaxial} orbits. Spherical orbits have conserved angular momentum, and in order to satisfy Eq. \ref{eq:lc} are necessarily contained at all times within the bottom dashed square region of the top panel of Fig. \ref{fig:lini_box}. Axisymmetric orbits, such as the saucer and tube orbits,  in turn have conservation of the z-component of angular momentum and are therefore necessarily contained in the $|L_z/L\_{crit}| < 1$ region of phase space, denoted by vertical dashed lines in the same figures. Finally, triaxial orbits (pyramids, boxes or chaotic orbits) can have arbitrarily small or large values in the $(L, L_z)$ plane as a result of a lack of an integral of motion in that plane. This simple classification scheme is inspired by Fig. 6 of \citet{Vasiliev_2015}, and we refer the reader to that study for more details.  We can therefore estimate orbital-type fractions of Pops. II and III using these criteria by looking at their total and $z$-component of angular momentum at previous times throughout the run.  We present these fractions in Table \ref{tab:tab_box}. We find that more than $76.2\%$ of Pop. II and III orbits can only originate in triaxial nuclei, because only  $23.8\%$  show consistent conservation of $L_z$ so that $|L_z/L\_{crit}| < 1$ is fulfilled throughout the run. This clearly shows that our centrophilic orbits are dominated by triaxial orbits by a factor of three, despite only a small deviation from axisymmetry of our system (Fig. \ref{fig:triax}).  
\begin{table}\centering
\caption{ Fraction of orbital families of centrophilic stars depending on the shape of the potential.}

\begin{tabular}{c | c  }
\hline
   Potential type &  Fraction of encounters $(\%)$  \\
 \hline
   \textit{Spherical} & 0.0   \\
  \hline
   \textit{Axisymmetric} & 23.8    \\
    \hline
   \textit{Triaxial} & 76.2   \\
     \hline \\

\end{tabular}
\tablefoot{ The classification is performed based on angular momentum changes of the stars throughout our run. Only stars in Populations II and III were considered.}
\label{tab:tab_box}
\end{table}

%%%%%%%%%%%%%%%%%%%%%%%%%%%%%%%%%%%%%%%%%%%%%%%%%%%55

\section{Discussion}
\label{sec:disc}
\subsection{Inclination of loss cone stars}
 \par   We find an overabundance of stars on prograde (co-rotating) orbits interacting strongly with the SMBH binary. This is expected, because stars on prograde  orbits have a higher chance of capture by the binary because of the larger orbital phase they have compared to the retrograde cases \citep{Wang2014, Milosavljevic2001}. This preference naturally results in an overabundance of counter-rotating orbits within close vicinity of the SMBH binary, as discovered by previous studies \citep{Meiron2010, Meiron2013}.
    We show that retrograde orbits can experience a change of sign in angular momentum during the interaction, in agreement with \citet{Wang2014}. These cases are also the most energetic events we measure, in agreement with \citet{Rasskazov2017} who assert that the total energy change of the star is proportional to its change in the component of angular momentum parallel to the binary. 
\subsection{Hybrid integration approach}
\par Our results demonstrate that the hybrid integration approach is very well suited for investigating close stellar interactions with an SMBH binary. The combination of the SCF force calculation for the outer regions with direct summation for inner particles of interest successfully resolves two of the biggest issues that typically plague N-body approaches. The first issue being the artificial enhancement of two-body relaxation effects that originates from insufficient mass resolution of N-body codes when compared to real galaxies. This issue has always pervaded N-body simulations and has previously cast doubt on measured hardening rates in N-body \citep{Vasiliev_2015}. The second issue resolved by the hybrid approach is the exceptionally high computational cost for simulations with $N\gtrsim10^6$. The proportionally small fraction of particles integrated in a direct way results in a factor of 16 speed-up over the pure N-body approach and enables simulations of  multi-million particle systems easily attainable even by smaller computing clusters with only a few computation nodes. 
\subsection{Merger timescale and numerical parameters}
\par     We note that the merging time of the SMBH binary reported in \citet{Khan2016} is underestimated by a factor of two as a result of numerical artifacts. Namely, we find that the rise in eccentricity of the binary in the original study is not due to stellar interactions, but a sign of insufficient integration accuracy in the black hole equations of motion, leading to an overestimation of the contribution of the PN terms and a premature PN plunge. We believe that this happens primarily because of two numerical parameters,  insufficient spatial resolution (gravitational softening) and the minimum black hole integration time-step. Our convergence tests show that at least the values of $\epsilon = 2\times10^{-4}$  $\mathrm{pc}$ for the softening and the minimum time-step of $\Delta t\_{min} = 10^{-5}$ $\mathrm{yr}$ are necessary to accurately integrate the SMBH binary evolution in this phase of the merger. Similarly, we find that this value of $\Delta t\_{min}$ becomes insufficient for the later part of the merger, when $t-t_0 > 4.65 \, \mathrm{Myr}$, which is why we stop the run at this point. Therefore,  this study focuses only on the phase of the merger when no PN effects are measurably present, because  even greater numerical accuracy would be necessary to accurately integrate the later phase of the merger when PN hardening becomes comparable to stellar hardening. This would result in a slowdown which would contribute to the already present and significant slowdown of the code in the PN regime. Because of this, accurate investigation of loss cone stars in this regime is inherently difficult and computationally intensive, even with the hybrid approach.  Nevertheless, we plan to explore this regime in an upcoming publication.
\par  We estimate the new merger timescale to be on the order of $\approx 20$ $\mathrm{Myr}$, which is still two orders of magnitude smaller than the Hubble time, thus avoiding the FPP. However, we note that the same limitations and uncertainties may apply as those that were discussed in \citet{Khan2016}.

\par    Finally, we also note that our simulation does not take into account the possibility of direct plunges of stars into the black hole horizon. As many of our stars are able to come exceedingly close to the binary ($<a\_{bh}$), it is natural  to assume that some of them will be lost in this way. This investigation is however beyond the scope of this study. Instead, we refer the reader to a few recent studies on 
tidal disruption events: \citet{Li2019, Lezhnin2019, Darbha2018}

\section{Conclusions}
\label{sec:conc}
\par In this paper, we present an N-body simulation of an unequal-mass SMBH binary system embedded in a dense, slightly triaxial, rotating stellar cusp. Our system originates from a high-redshift major galactic merger in an \textit{ab initio} cosmological simulation, as described in \citet{Khan2016}. With a particle number of $6 \times 10^6$, we simulate the hardening phase of the SMBH binary merger and explore in detail the properties of the stellar particles which experience energetic interactions with the black holes. We use the $\varphi$-GRAPE-hybrid code, which combines direct integration with the collisionless SCF integration method to considerably reduce computational cost and spurious relaxation effects and enable exploration of the orbital parameters of stars in the loss cone with high accuracy. We can summarize our main conclusions as follows: 
\begin{itemize}
\vspace{6pt}
\item 
To a very high degree of accuracy, we are able to identify the exact stars that contribute to the SMBH binary hardening. Our energy balance plots (see Fig. \ref{fig:gap}) show that throughout the run,   the cumulative energy changes of stars within the high-energy tail correspond almost exactly to the overall SMBH binary orbital energy change for the same time intervals. This demonstrates beyond any doubt that stellar hardening is the main driver of SMBH binary energy loss in our system, and that other possible effects of energy loss can be neglected. This further implies that in gas-poor systems, proper treatment of stellar scattering interactions, either via simple analytic recipes derived on scattering experiments \citep[e.g.,][]{Sesana2015, Sesana2010} or via direct summation, can be sufficient to properly characterize the evolution of SMBH binaries. 

\item
We distinguish three populations of stellar encounters based on their apocenter distributions (see Fig. \ref{fig:rap_log}). Population I originates close to the binary, contributes the least to hardening, and is a consequence of the Brownian random motion of the SMBH binary around the center.  Population II originates from outside the influence radius, with apocenters $r\_{ap}\approx5R\_{infl}$ and is characterized by fast, single hyperbolic and parabolic interactions. This population has by far the highest effect on the SMBH hardening, because it is the most populous. Population III has similar properties as Pop. II, but becomes bound to the binary and is put on eccentric orbits.
\vspace{6pt}

\item We identify Pops. II and III as the centrophilic orbits responsible for refilling the loss cone. We analyze their angular momentum changes and estimate that $76.2\%$ of these orbits can originate only in a triaxial potential, strongly supporting the hypothesis that even slight triaxiality is crucial for the efficient hardening of the SMBH binary that we measure, in agreement with the findings of \citet{Vasiliev_2014, Bortolas2018BHB, Khan2018b}.
\vspace{6pt}

\item Most of the energetic interactions show prograde rotation with the SMBH binary, as well as the overall system.  However,  the retrograde interactions correspond to the most energetic interactions that we measure and can result in a change of sign in angular momentum. Because of this, we find that despite being significantly lower in number, retrograde orbits make up $45\%$ of the total energy exchange.  
\vspace{6pt}

\end{itemize}

\begin{acknowledgements}
\par We would like to thank the anonymous reviewer for the detailed review and for their comments and suggestions that greatly improved the manuscript. BA thanks Manuel Arca Sedda for his comments on the manuscript,  as well as many fruitful discussions and comments. BA also acknowledges Rainer Spurzem and Christian Fendt for helpful discussions and feedback, as well as the support from the International Max Planck Research School (IMPRS) at the University of Heidelberg. This work was funded by a \emph{“Landesgraduiertenstipendium"} of the University of Heidelberg. It is also partly funded by the Volkswagen Foundation under the Trilateral Collaboration Scheme (Russia, Ukraine, Germany) project titled ("Accretion Processes in Galactic Nuclei") (funding for personnel and international collaboration exchanges).  The authors gratefully acknowledge the Gauss Centre for Supercomputing e.V. (www.gauss-centre.eu) for funding this project by providing computing time through the John von Neumann Institute for Computing (NIC) on the GCS Supercomputer JUWELS at Jülich Supercomputing Centre (JSC). The  authors  also  acknowledge  support  by  the  state  of Baden-W\"urttemberg through bwHPC.
PB acknowledges support by the Chinese Academy of Sciences
through the Silk Road Project at NAOC, the President’s
International Fellowship (PIFI) for Visiting Scientists
program of CAS, the National Science Foundation of China
under grant No. 11673032.

This work was supported by the Deutsche Forschungsgemeinschaft
(DFG, German Research Foundation) – Project-ID 138713538 –
SFB 881 (‘The Milky Way System’), by the Volkswagen Foundation
under the Trilateral Partnerships grants No. 90411 and 97778.

The work of PB was supported under the special program of the NRF 
of Ukraine ‘Leading and Young Scientists Research Support’ - 
"Astrophysical Relativistic Galactic Objects (ARGO): life cycle of 
active nucleus",  No. 2020.02/0346.

\end{acknowledgements}

%------------------------

\bibliographystyle{aa}
\bibliography{mybibfile}

\end{document}